\documentclass[prl,twocolumn,aps,letterpaper, preprintnumbers,superscriptaddress]{revtex4}
\usepackage{hyperref}
\usepackage{verbatim}
\usepackage{amsmath}
\usepackage{latexsym}
\usepackage{revsymb}
\usepackage{yfonts}
\usepackage{ifthen}
\usepackage{tcolorbox}
\usepackage{natbib}
\usepackage{amsfonts}
\usepackage{amsmath}
\usepackage{amssymb}
\usepackage{amsthm}
\usepackage{graphicx}
\usepackage{bm}
\usepackage{bbm}
\usepackage{epsfig,color,amssymb}
\usepackage{subfigure}
\usepackage{amsfonts}
\usepackage{amscd}
\usepackage{amsmath}
\usepackage{multirow}
\usepackage{chemarrow}
\usepackage{dcolumn}
\usepackage{bm}
\usepackage{graphicx}
\usepackage{enumerate}
\usepackage{epsfig}
\usepackage{subfigure}
\usepackage{xcolor}
\usepackage{multirow}
\usepackage{ulem}
\usepackage{braket}
\usepackage{comment}
\usepackage{enumitem}
\usepackage{amsthm}

\renewcommand{\emph}[1]{\textit{#1}}

\begin{document}

\title{Optical skyrmions and other topological quasiparticles of light}

\author{Yijie Shen}\email{y.shen@soton.ac.uk}
\affiliation{Division of Physics and Applied Physics, School of Physical and Mathematical Sciences, Nanyang Technological University, Singapore, Singapore}\affiliation{Centre for Disruptive Photonic Technologies, The Photonics Institute, Nanyang Technological University, Singapore, Singapore}\affiliation{Optoelectronics Research Centre, University of Southampton, Southampton SO17 1BJ, United Kingdom}
\author{Qiang Zhang}
\author{Peng Shi}
\author{Luping Du}
\affiliation{Nanophotonics Research Center, Shenzhen Key Laboratory of Micro-Scale Optical Information Technology \& Institute of Microscale Optoelectronics, Shenzhen University, Shenzhen 518060, China}
\author{Xiaocong Yuan}\email{xcyuan@szu.edu.cn}
\affiliation{Nanophotonics Research Center, Shenzhen Key Laboratory of Micro-Scale Optical Information Technology \& Institute of Microscale Optoelectronics, Shenzhen University, Shenzhen 518060, China}
\author{Anatoly V. Zayats}\email{a.zayats@kcl.ac.uk}
\affiliation{Department of Physics and London Centre for Nanotechnology, King’s College London, London WC2R 2LS, United Kingdom}


\begin{abstract}
\noindent\textbf{Skyrmions are topologically stable quasiparticles that have been predicted and demonstrated in quantum fields, solid-state physics and magnetic materials, but only recently observed in electromagnetic fields. Here we review the recent advances in optical skyrmions within a unified topological framework. Starting from fundamental theories and classification of skyrmionic states, we describe generation and topological control of different kinds of skyrmions in evanescent, structured and spatiotemporal optical fields. We further highlight generalized classes of optical topological quasiparticles beyond skyrmions and outline the emerging applications, future trends and open challenges. A complex vectorial field structure of optical quasiparticles with versatile topological characteristics emerges as an important feature in modern spin optics, imaging, metrology, optical forces, structured light, and topological and quantum technologies.}
\end{abstract}

\maketitle

\noindent
In 1961, British particle physicist Tony Skyrme proposed a model of localized solitons to represent the topologically protected structures of a general class of nucleons~\cite{skyrme1991non}. He demonstrated that the pion structures have the similar properties as baryons (protons and neutrons) and proposed a unified expression parametrized by a topological number (baryon number) with a meson regarded as a special case when a topological number is zero~\cite{skyrme1962unified}. This came as a surprise, because protons and neutrons are fermions, whereas pions are bosons, and the Skyrme’s topological theory provided a unified description~\cite{zahed1986skyrme}. The quasiparticles emerging in this mathematical model were named as skyrmions. Since then, skyrmions—the generalized topological quasiparticles—have been considered in many contexts, including nucleons~\cite{naya2018skyrmions,halcrow2020attractive}, 
Bose-Einstein condensates~\cite{al2001skyrmions}, liquid crystals~\cite{duzgun2021skyrmion}, magnetic materials~\cite{liu2016skyrmions,tokura2020magnetic,bogdanov2020physical,fert2017magnetic,lima2022spin,han2022high}, and recently twistronics~\cite{khalaf2021charged,kwan2022skyrmions}. The most known type of skyrmions is the magnetic skyrmions in chiral magnets formed by a magnetization texture~\cite{muhlbauer2009skyrmion}, which are readily observable in real space~\cite{yu2010real}. Magnetic skyrmions opened new avenues in high-density data storage and low-energy magnetic memory~\cite{fert2017magnetic,lima2022spin,han2022high}.

Very recently, skyrmionic structures were observed in optics~\cite{tsesses2018optical,du2019deep}. A lattice of skyrmion-shaped distributions of the evanescent electric field, oscillating in time but preserving their topology, was realised in the especially shaped surface-plasmon polariton (SPP) resonator~\cite{tsesses2018optical}. In parallel, a stable, individual skyrmions formed by the spin distribution of a SPP field in the presence of spin-orbit coupling was demonstrated~\cite{du2019deep}. These observations of optical fields emulating skyrmionic topological quasiparticle properties opened up a new chapter in modern optics, attracting continuously growing body of research on skyrmionic properties of electromagnetic fields, enabling new approaches to achieving and controlling topological properties of light in space and time, structured light, and light–matter interactions~\cite{rivera2020light}. The related applications span from spin-optics, imaging and metrology to optical forces, structured light and topological and quantum technologies.

In this Review, we outline recent advances in optical skyrmions, present a unified framework for classification of their topological states, and discuss approaches for generation and control of skyrmions in various realizations. We further discuss topological optical field description towards generalized topological quasiparticles beyond skyrmions, such as merons and hopfions.  Finally, we overview potential applications of topological light fields, from microscopy, communications and metrology to enhanced light-matter interactions, and look at the open challenges and prospects of this fascinating field.\\[4pt]

\noindent
\textbf{\large Classification of topological quasiparticles}\\[2pt]
\noindent
\\
Physically, a skyrmion is a quasiparticle carrying a topologically protected vector texture. Geometrically, it is a topologically stable 3D continuous vector field configuration, restricted within a confined space, in an otherwise homogeneous infinite medium, describing a topological soliton solution~\cite{manton2004topological}. In the case of magnetic systems, magnetic skyrmions almost exclusively been considered in a compact 2D plane, and often called ``baby skyrmions''~\cite{tokura2020magnetic,bogdanov2020physical,fert2017magnetic}. Here, we follow the convention for magnetic skyrmions and refer to such a 2D skyrmion as a skyrmion. 

\begin{figure*}[t!]
	\centering
	\includegraphics[width=0.87\linewidth]{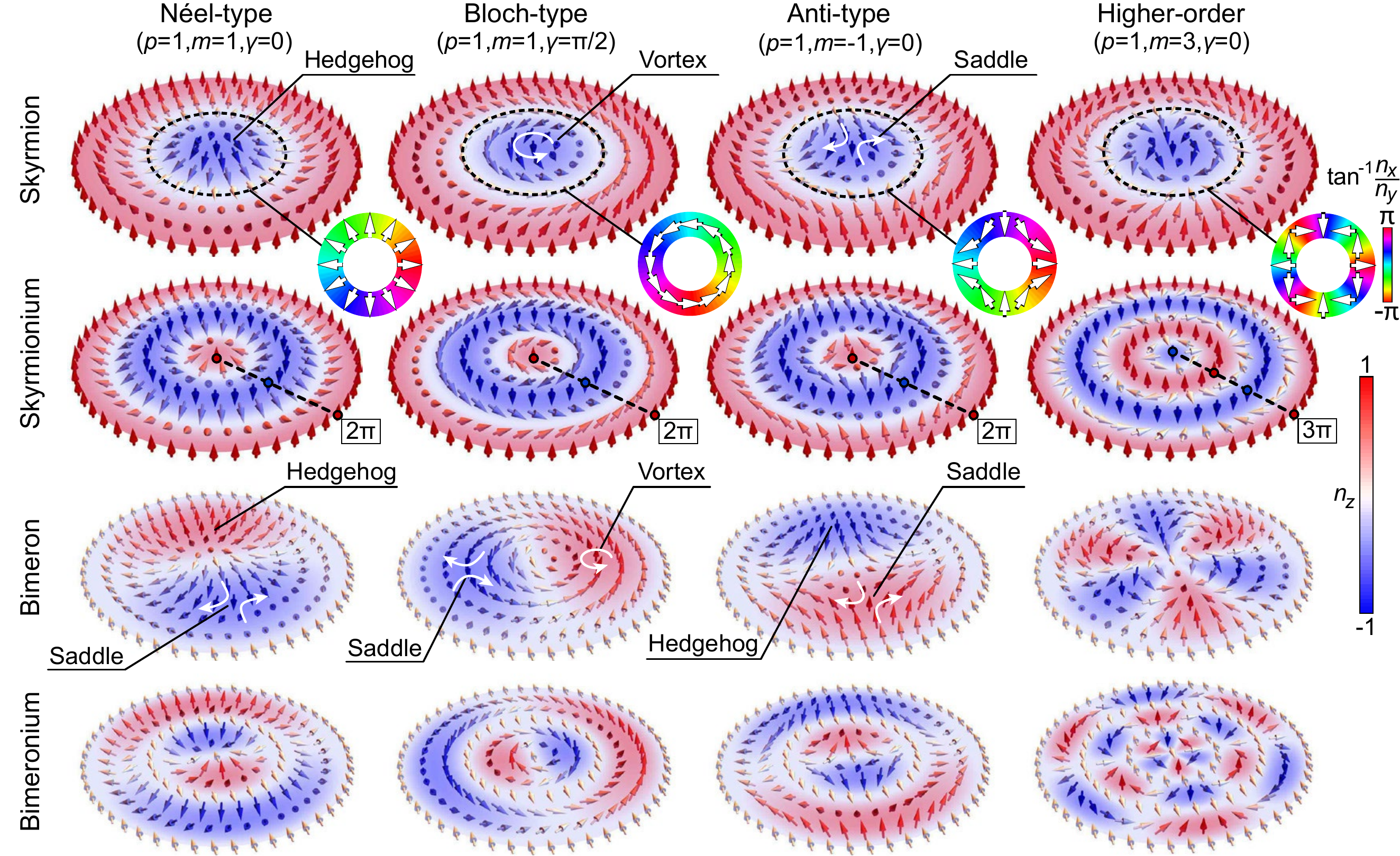}
	\caption{\textbf{Topological quasiparticle classification.} (top to bottom) Quasiparticles with classifications of skyrmion, skyrmionium, bimeron, and bimeronium. In each class, the quasiparticles are also classified by different topological textures as (left to right) N\'eel type, Bloch type, anti-type, and higher-order (third-order) skyrmions. Topological numbers of polarity, vorticity and helicity for each case are also shown. For skyrmions (top row), dashed lines indicate the position at which the orientation of the transverse vector components is shown in the inserts. For skyrmionium (second row), dashed lines highlight the radial $k\pi$-twist.} 
	\label{f1}
\end{figure*}

\begin{widetext}
\noindent\fbox{\begin{minipage}{50em}
\textbf{\textbf{Box 1 $|$ Topological mapping of a vector field and characteristics.}}
A topological texture can be constructed by a vector field fulfilling a mapping from a parameter space to real space, e.g., from 4D to 3D real space or from 3D to 2D real space, similar to a stereographic projection~\cite{hopf1964abbildungen}. In the former case, the mapping is from a hypersphere $\mathbb{S}^3$ formed by 4D vectors $|\bm{\zeta}\rangle=(\chi_4+\text{i}\chi_3,\chi_1+\text{i}\chi_2)^{\text{T}}$, represented by four real variables, usually in the SU(2) group (two complex variables and $\chi_1^2+\chi_2^2+\chi_3^2+\chi_4^2=1$). The projection can be achieved by $\mathbf{S}=\langle\bm{\zeta}|\bm{\sigma}|\bm{\zeta}\rangle$, where 
$\mathbf{S}=[s_1,s_2,s_3]\in\mathbb{R}^3$ is a 3D space with three real variables, i.e. $s_i=\langle\bm{\zeta}|\bm{\sigma}_i|\bm{\zeta}
$ ($i=1,2,3$), which form a 2-sphere $\mathbb{S}^2$ ($s_1^2+s_2^2+s_3^2=1$), and $\bm{\sigma}=[\bm{\sigma}_1,\bm{\sigma}_2,\bm{\sigma}_3]$ refers to the three Pauli matrices. In the Skyrme model, only a reduced 2D space of a 4D vector is considered, e.g., $(\chi_1,\chi_2)=(x,y)/r$, where $r^2=x^2+y^2$, and the vector textures localized in a 2D $x$-$y$ plane can be constructed, i.e., skyrmions and merons (Figs.~\ref{f1}\textbf{a-e}). In the generalized Skyrme-Faddeev model~\cite{faddeev1976some}, a 4D vector is projected onto 3D Cartesian space by $(\chi_1,\chi_2,\chi_3,\chi_4)=\left(\frac{x}{r}\sin f,\frac{y}{r}\sin f,\frac{z}{r}\sin f, \cos f\right)$, where $r^2=x^2+y^2+z^2$ and $f$ is the phase parameter. In this case, the vector textures in 3D space $(x,y,z)$ are termed hopfions. The isospin lines of a hopfion form a complex configuration, a Hopf fibration (Fig.~\ref{f1}\textbf{f}), on a surface of nested tori (Fig.~\ref{f1}\textbf{g}). 2D skyrmions can be seen as a special case of a 3D hopfion. This calls the mapping from a 3D parametric space (spin space) to a 2D real space by setting $z=0$ and $f=0$, resulting in a 2D baby skyrmion (Fig.~\ref{f1}\textbf{a,e}), which can be found in the subspace of a hopfion. The mapping may also result in more complex textures, such as the links of elementary hopfions (Fig.~\ref{f1}\textbf{h}). 
\\
The vector field which forms a 2D skyrmion, 
$\mathbf{S}(x,y)=[n_x(x,y),n_y(x,y),n_z(x,y)]$,
can be represented by the vector  distribution  unwrapped  from  the  vectors  on  a sphere parametrized by longitude, $\alpha$, and latitude, $\beta$, angles (Fig.~\ref{fbox1}). The topology of a skyrmionic configuration can be characterized by the skyrmion number defined as~\cite{gobel2021beyond,nagaosa2013topological}
\begin{equation}
s=\frac{1}{4\pi }\iint_\sigma{\mathbf{S}\cdot \left( \frac{\partial \mathbf{S}}{\partial x}\times \frac{\partial \mathbf{S}}{\partial y} \right)}\text{d}x\text{d}y
\end{equation}
where $\mathbf{S}(x,y)= \mathbf{S}(r\cos\phi, r\sin\phi)$ describes  the vector field which constructs a quasiparticle and $\sigma$ is the region where it is confined. The skyrmion number can be phenomenologically understood as a number of times the vectors wrap around a unit sphere. In an example of a skyrmion with $s=1$, the unit sphere mapping is shown in Fig.~\ref{fbox1}\textbf{a}, where the sets of vectors from the skyrmion center to its boundary are mapped onto a sphere from the south pole to the north pole. However, a skyrmion number cannot uniquely determine the spin texture. Based on the mapping, the quasiparticle vectors can be written as $\mathbf{S}=(\cos\alpha (\phi )\sin\beta (r),\sin\alpha (\phi )\sin\beta (r),\cos\beta (r))$, and the skyrmion number can be separated into additional topological numbers: 
\begin{align}
s=\frac{1}{4\pi }\int_{0}^{r_\sigma}{\text{d}r}\int_{0}^{2\pi }{\text{d}\phi }\frac{\text{d}\beta (r)}{\text{d}r}\frac{\text{d}\alpha (\phi )}{\text{d}\phi }\sin \beta (r)
=\frac{1}{4\pi }[\cos\beta (r)]_{r=0}^{r=r_\sigma}[\alpha (\phi )]_{\phi =0}^{\phi =2\pi }=p\cdot m
\end{align}
with the integer $p=\frac{1}{2}[\cos\beta (r)]_{r=0}^{r=r_\sigma}$ defining the polarity, i.e., the vector direction down (up) at center $r=0$ and up (down) at boundary $r\to r_\sigma$ for $p=1$ ($p=-1$), and the integer $m=\frac{1}{2\pi }[\alpha (\phi )]_{\phi =0}^{\phi =2\pi }$ defining the vorticity, which controls the distribution of the transverse field components. For distinguishing different helical vortexes, an initial phase $\gamma$ should be added, $\alpha (\phi)=m\phi +\gamma$. 
\\
\\
By unwrapping a parametric sphere in different ways, various quasiparticles with different topological properties can be obtained. A skyrmion with $s=1$ but with the vector structure arranged along tangent direction (Fig.~\ref{fbox1}\textbf{b}) represents a vortex (Bloch-type)  texture with $\gamma=\pi/2$, which is different from a hedgehog (N\'eel-type) texture with $\gamma=0$ (Fig.~\ref{fbox1}\textbf{a}). A skyrmion with opposite vorticity $m=-1$, termed anti-skyrmion, has a very different to conventional skyrmion saddle texture (Fig.~\ref{fbox1}\textbf{c}). A skyrmion can be divided into two merons with a half skyrmion number, each corresponding to a half unwrapping from the parameter sphere. By changing the unwrapping style of the vectors around a sphere, other types of quasiparticles can be obtained. If the sphere is unwrapped not from the pole but from a point on the equator, a bimeron quasiparticle is obtained (Fig.~\ref{fbox1}\textbf{d}). These classification of topological textures are purely mathematical in origin, and for optical realisations, physical electromagnetic vector fields should be used to describe optical quasiparticles.

\end{minipage}}

\begin{figure}[!h]
	\centering
	\includegraphics[width=0.96\linewidth]{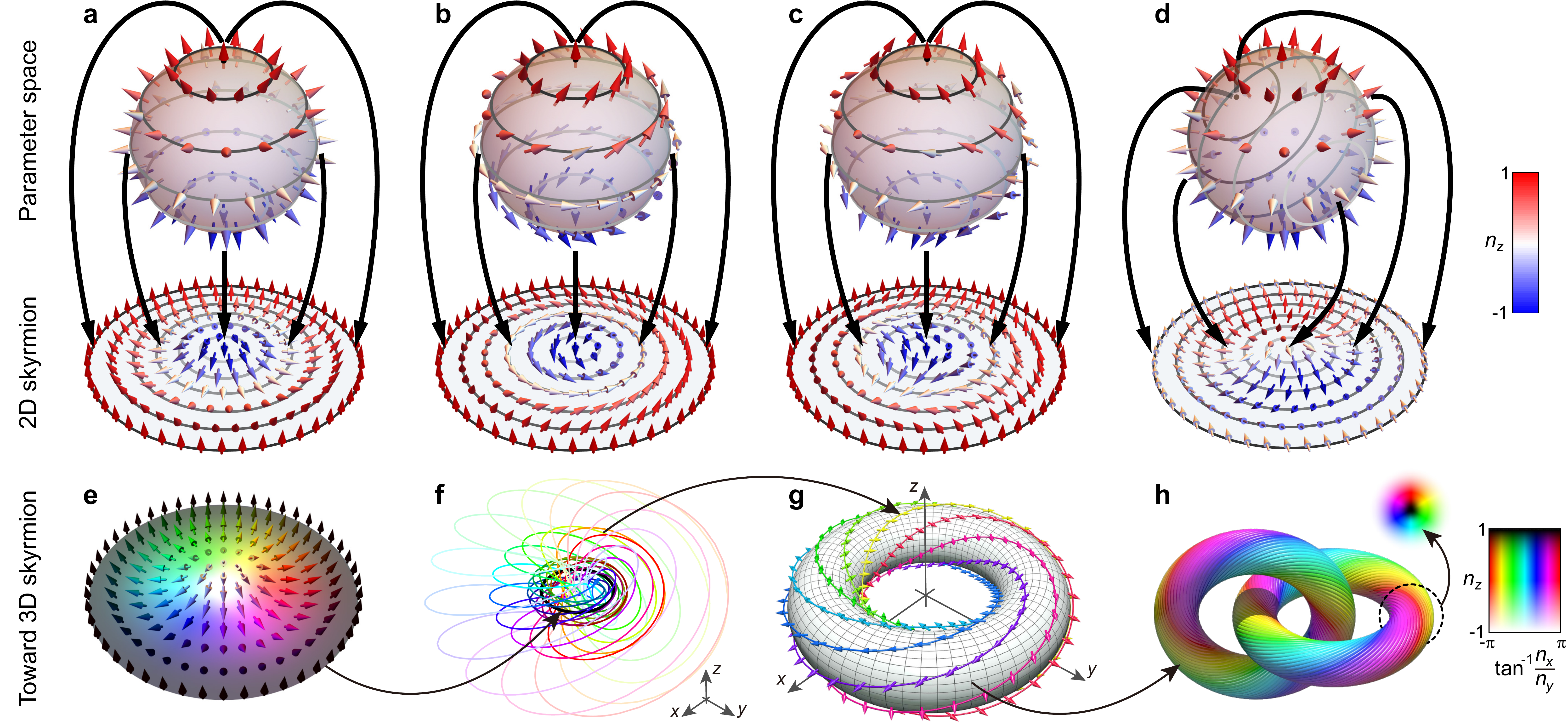}
	\caption{\textbf{Parametric sphere representations of quasiparticles (figure for Box~1).} \textbf{a}, A N\'eel-type skyrmion ($s=1$, $\gamma=0$) with a hedgehog-like texture can be obtained by mapping the vectors on a sphere onto a confined plane starting at the pole. \textbf{b}, A Bloch-type skyrmion ($s=1$,$\gamma=1/2$) with vortex texture. \textbf{c}, An anti-skyrmion ($s=-1$) with an opposite skyrmion number to N\'eel- or Bloch skyrmions. \textbf{d}, If the initial (central) point of the unwrapping is chosen at the equator, a bimeron ($s=1$) is obtained. \textbf{e,f}, In addition to the 2D skyrmion (\textbf{e}), other vector textures can be generated in 3D, such as a hopfion fulfilling a Hopf fibration (\textbf{f}), where a 2D skyrmion (\textbf{e}) can found in its subspace. \textbf{g}, A hopfion refers to a 3D spin texture with the isospin lines on a  surface of nested tori (one torus is shown). \textbf{h}, A hopfion link (insert shows a crossecton of a hopfion.}
\end{figure}

\end{widetext}

The vector field which forms a 2D skyrmion can be obtained as the vector spatial distribution unwrapped from the vectors on a parametric sphere (Box~1). Various topological textures of skyrmions can be achieved by appropriately arranging vectors on the parametric sphere or changing the way how they are unwrapped. The topological property of a skyrmionic configuration is characterized by a skyrmion number $s$, as well as additional topological numbers, such as polarity $p$, vorticity $m$, and  helicity $\gamma$, as defined in Box~1. 

These topological numbers determine classification of quasiparticle types. In the case of $m=1$, the skyrmions of $\gamma=0$ or $\gamma=\pi$ are classified as \textit{N\'eel-type} and exhibit a hedgehog texture around the quasiparticle~\cite{kezsmarki2015neel}. For $\gamma=\pm\pi/2$, the skyrmions are classified as \textit{Bloch-type} with a vortex texture around the center~\cite{milde2013unwinding}. In the case of $m=-1$, the topology is defined as anti-skyrmion with a saddle (anti-vortex) texture~\cite{nayak2017magnetic}. Anti-skyrmions cannot be classified by anti-N\'eel or anti-Bloch types due to the intrinsic symmetry of a saddle structure, which is only wholly rotate itself by tuning helicity. Geometrically, the distinction of an anti-skyrmion is a counter-clockwise vector distribution along the perimeter encircling a skyrmion center; the vectors of an anti-skyrmion wrap around the center clockwise ($-2\pi$), while the vectors of a skyrmion wrap around the centre counter-clockwise ($2\pi$) (Fig.~\ref{f1}). In the case of large values of vorticity associated with larger value of a skyrmion number, the skyrmions are designated as higher-order skyrmions~\cite{zhang2017direct} (Fig.~\ref{f1}). Physical realisation of different unwrapping styles is related to a particular system properties. For example, the N\'eel and Bloch types of magnetic skyrmions are determined by the behavior of domain walls with different magnetic anisotropy. For optical skyrmions, symmetry breaking is one of the factors that defines an unwrapping style.

Skyrmions have many generalized forms in the quasiparticle family, a typical example of which is skyrmionium, a coupled state connecting two skyrmions with opposite polarities and resulting into skyrmion number of $s=1-1=0$~\cite{zhang2016control}. The skyrmionium has zero total skyrmion number and a radial $2\pi$-twist structure (``$k\pi$-twist'', $k=1,2,3,\cdots$, refers to the a $k\pi$ angle with which a vector twists repeatedly upon the evolution along radial direction), which is a basis for many anomalous physical effects in magnetics~\cite{kolesnikov2018skyrmionium}.The  topological classification of N\'eel-, Bloch-, and anti-types of skyrmions is also applicable for skyrmionium (Fig.~\ref{f1}). A natural generalization of skyrionium is a $k\pi$-skyrmion with the raidal multi-($k\pi$-)twist structure~\cite{song2019field}, also known as a target skyrmion~\cite{zheng2017direct}, which replaces the radial $2\pi$-twist structure with $k\pi$-twist structure and controls the skyrmion number between $0$ and $1$ (or $-1$).

Another important kind of topological quasiparticles is merons and bimerons. A bimeron can similarly be described by a unit sphere mapping but with a more general wrapping style (Box~1). In contrast to the skyrmion representation with a fixed spiny sphere, a bimeron can be obtained by switching the coordinates of longitude and latitude or tilting the sphere. Based on the generalized topological transformation, the bimeron has the texture composed by two merons with opposite polarities~\cite{jani2021antiferromagnetic}. The above topological classification is also applicable to merons and bimerons~\cite{desplat2019paths,bera2019theory} (Fig.~\ref{f1}). (Please note that in some initial papers, anti-type of (bi)meron was introduced due to its negative polarity~\cite{yu2018transformation}, which is different from an anti-(bi)meron notion introduced here). Recently, a new concept of bimeronium was proposed as an extension of bimeron~\cite{zhang2021frustrated}, similar to skyrmionium evolved from skyrmion, with similar topological classification. The described above topological classifications can be extended to considerations of any quasiparticle, for example, the $k\pi$-bimeron with additional radial multi-twist structure in bimeronium (fourth plot in the fourth row of Fig.~\ref{f1}), which yet to be found experimentally. The described methodology can guide future discoveries of new types of topological quasiparticles in physics in general and photonics in particular. 
\\

\noindent
\textbf{\large Optical skyrmions}\\[2pt]
\noindent



\begin{figure}[!h]
	\centering
	\includegraphics[width=\linewidth]{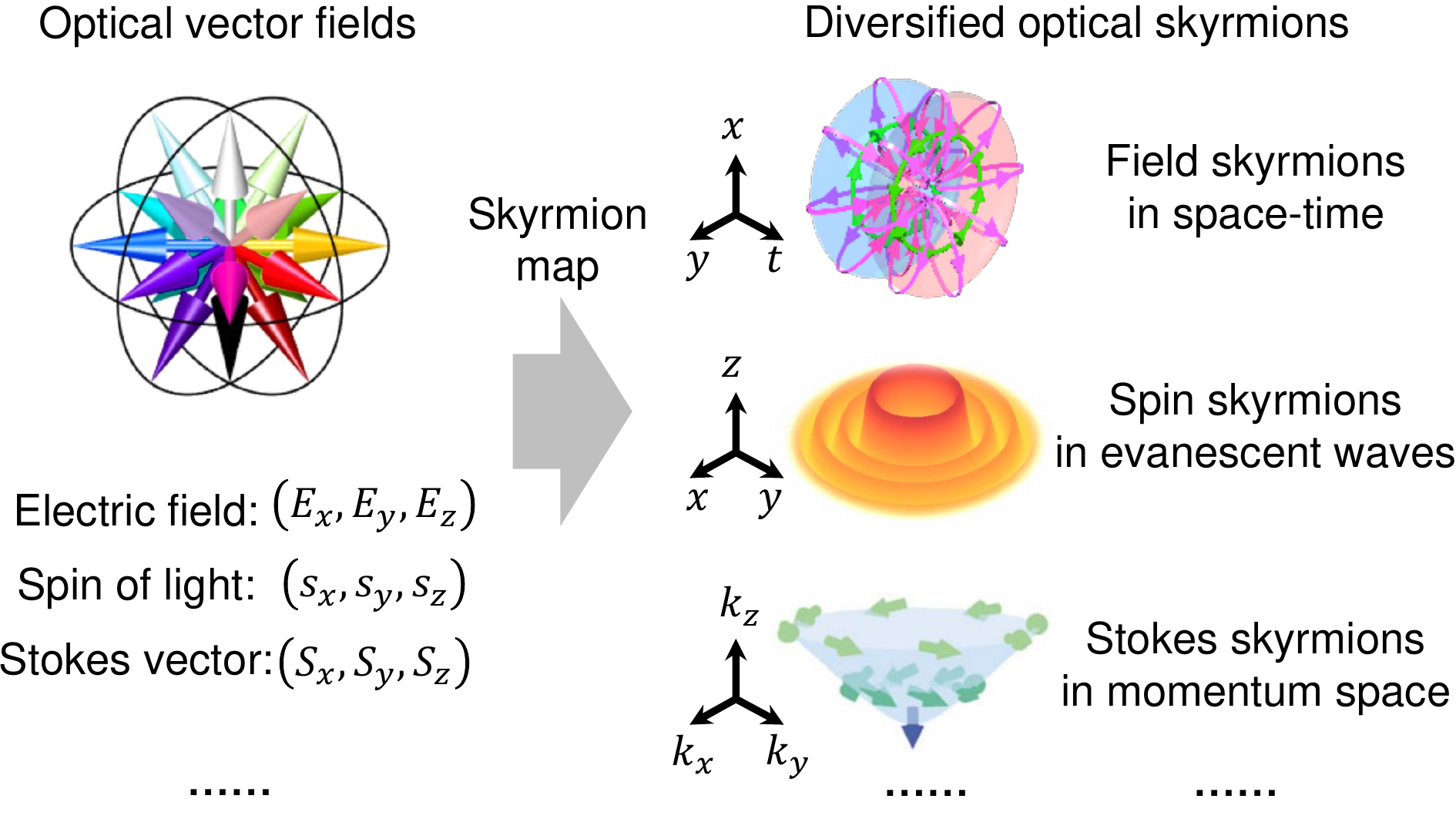}
	\caption{\textbf{Generation of optical skyrmions.} Choosing the vector field appropriate to optical configuration, such as electric field, spin or Stokes vectors and physical realisation for its mapping in space or space-time, the optical textures satisfying skyrmion mapping can be obtained. Topological protection is provided by underlying properties of the optical field and physical domain where a quasiparticle is formed.}
	\label{fbox3}
\end{figure}

Optical skyrmions and other quasiparticles can be constructed with various 3D vector fields, which in optics can be, for example, the electric field $(E_x,E_y,E_z)$, spin angular momentum $(s_x, s_y, s_z)$, polarization Stokes vector $(S_x, S_y, S_z)$, pseudospin $(k_x,k_y,k_z)$, etc. 2D surface or 3D space or 4D space-time physical representations can be used to obtain various kinds of optical quasiparticles, fulfilling the appropriate mapping (Fig.~\ref{fbox3}). 
\\[4pt]


\noindent
\textbf{Skyrmions in evanescent fields}\\[2pt]
The main challenge in achieving optical skyrmions is the ability to twist the optical field vectors in 3D in on-demand structured fields~\cite{forbes2021structured}, which requires to break the prevailing paradigm to treat electromagnetic field as a transverse field. Longitudinal components naturally appear due to the evanescent field present in guided modes of planar or confined waveguides. Therefore, 2D guided electromagnetic modes, such as surface plasmon polaritons or other planar waveguides are obvious candidates for realisation of 2D topological structures. 
\\[4pt]



\begin{figure*}[ht!]
	\centering
	\includegraphics[width=\linewidth]{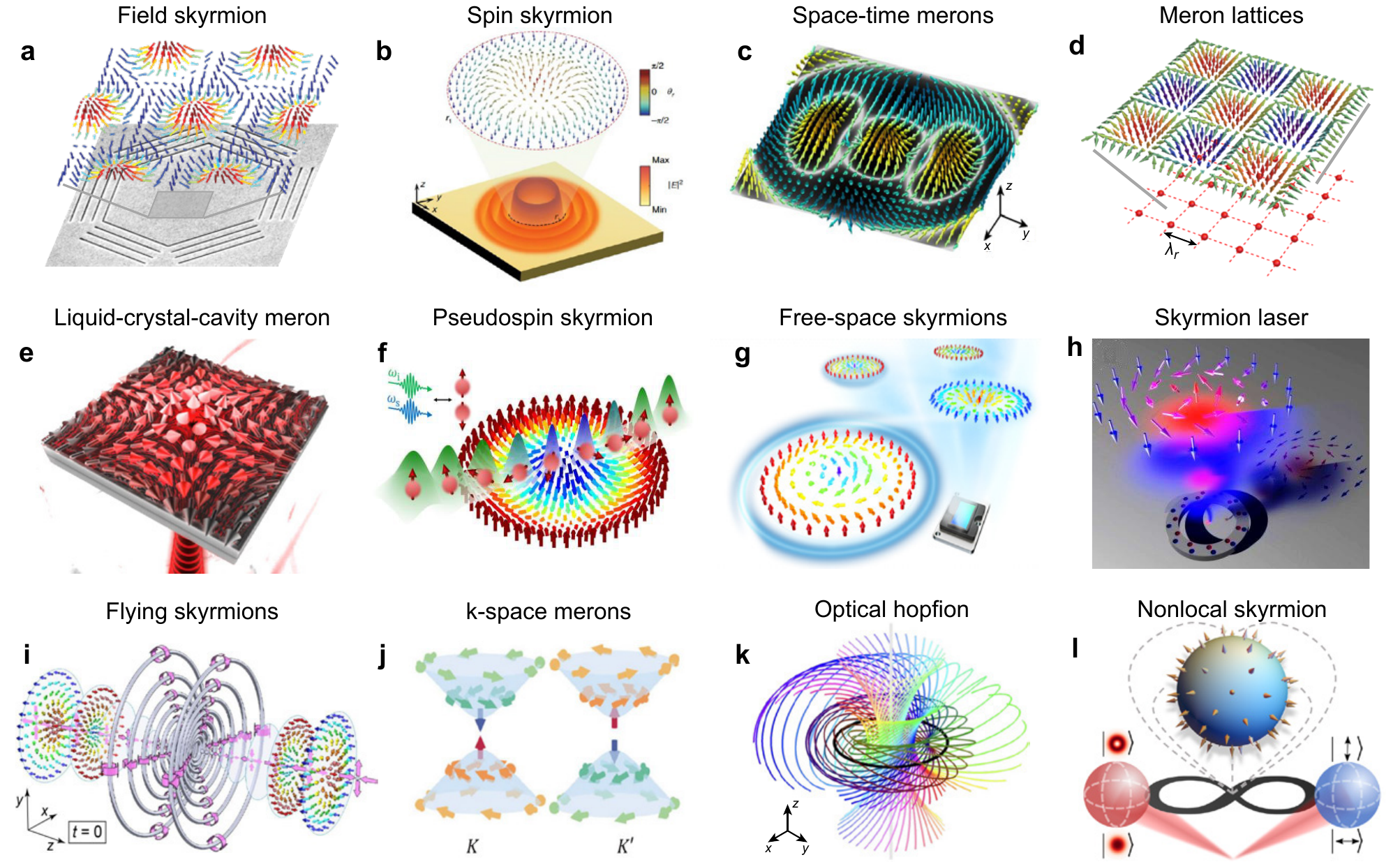}
	\caption{\textbf{Optical skyrmions.} \textbf{a}, Optical field-skyrmion lattices generated in the evanescent electric field of a SPP wave~\cite{tsesses2018optical}. \textbf{b}, An optical spin-skyrmion generated in a SPP wave with orbital angular momentum~\cite{du2019deep}. \textbf{c}, An optical spin-skyrmion in a SPP wave generated with an Archimedean geometric structure ~\cite{dai2020plasmonic}. \textbf{d}, Optical spin-meron lattices generated in the SPP field of four-fold symmetry  in the presence of OAM~\cite{lei2021photonic}. \textbf{e}, An optical meron generated in a liquid-crystal-filled cavity~\cite{krol2021observation}. \textbf{f}, An optical skyrmion constructed by pseudospins of a nonlinear photonic crystal~\cite{karnieli2021emulating}.  \textbf{g}, Optical Stokes skyrmions  with topological textures between N\'eel-, Bloch-, and anti-skyrmions~\cite{shen2021generation}. \textbf{h}, An optical Stokes skyrmion in a paraxial full Poincar\'e beam emitted by a microlaser~\cite{lin2021microcavity}. \textbf{i}, The supertoroidal light pulse carrying multiple field-skyrmions propagating in free space~\cite{shen2021supertoroidal}. \textbf{j}, The skyrmion textures observed in the momentum space of a photonic crystal slab~\cite{guo2020meron}. 
 \textbf{k}, 
 The 3D polarization texture of the vector beam representing an optical hopfion, where each Hopf fiber is constructed by a trajectory of a certain polarization ellipse (The hue colors depict different ellipse azimuths and dark to bright progression corresponds to the polarisation variation from left- to right-handed circular state)~\cite{sugic2021particle}.
 \textbf{l}, A nonlocal optical skyrmion as a quantum entangled state of two photons with hybrid degrees of freedom, spatial mode and polarization, fulfilling a skyrmion mapping in the correlation function of the two entangled photons~\cite{ornelas2022non}.} 
	\label{f2}
\end{figure*}

\noindent\textbf{Field skyrmions.} 
Optical field-skyrmion lattice was first observed in the evanescent electric field of the SPP waves interfering in a resonator of a six-fold symmetry~\cite{tsesses2018optical}. 
In such a resonator, the evanescent electric field of the SPP constructs skyrmionic structures, with out-of-plane field $E_z=E_0e^{-k_zz}\sum_{\theta=0,\pm\frac{\pi}{3}}\cos[k_r(x\cos\theta+y\sin\theta)]$ and in-plane field $(E_x,E_y)=-\frac{k_z}{k_r^2}(\frac{\partial E_z}{\partial x},\frac{\partial E_z}{\partial y})$, where $k_z$ and $k_r$ is the out-of-plane and in-plane wavevectors, respectively. 
Each pair of gratings on opposite sides of the hexagon supports a standing wave and the skyrmion textures are formed by superposition of the electric fields of the standing waves along variant spatial directions
(Fig.~\ref{f2}\textbf{a}). 
Dynamics of such skyrmionic SPP standing wave can be observed using photoemission electron microscopy~\cite{davis2020ultrafast}. 
Spoof SPP, the surface electromagnetic waves that propagate along planar interfaces with sign-changing permittivities were also reported to generate skyrmion-like structures for the fields in a resonator~\cite{deng2021observation,yang2021symmetry}, but their topological protection needs to be further investigated. 
SPP field-skyrmions have the topological texture limited to N\'eel-type topology.  While optical field-skyrmions which can be switched between N\'eel- and Bloch-types were reported~\cite{bai2020dynamic}, the existence of Bloch-type skyrmions was promptly disproved~\cite{meiler2020dynamic,bai2020dynamicr}. Photonic field-skyrmions, which are dynamic field textures with the electric field oscillating in time, exist only in a six-fold lattice. The more general topological textures in field-skyrmions are still to be found.
\\[4pt]

\noindent\textbf{Spin skyrmions.}
In contrast to time-varying field-skyrmion lattices, static optical skyrmions can be constructed from the spin vectors (spin angular momentum, SAM) of the evanescent electromagnetic field, with out-of-plane and radial components of $S_z\propto \frac{k_r^3}{r}J(k_rr)J'(k_rr)e^{-2k_zz}$ and $S_r\propto \frac{k_r^2k_z}{r}J^2(k_rr)e^{-2k_zz}$, where $J$ denotes the first-order Bessel function. The same as the field-skyrmions, optical spin-skyrmions also allow only N\'eel-type textures in the isotropic medium (Fig.~\ref{f2}\textbf{b})~\cite{du2019deep}. 
They exist in the evanescent fields carrying orbital angular momentum (OAM), and can be realised by exciting a waveguided mode with a light having OAM. Such optical skyrmions are most closely related to their magnetic counterparts and can exist as individual isolated quasiparticles~\cite{du2019deep} as well as form hexagonal lattices~\cite{lei2021photonic}. 

The radial $k\pi$-twist structures of the spin-skyrmion possess deep-subwavelength features, opening a new pathway for super-resolution imaging and microscopy~\cite{du2019deep}. On the other hand, exploiting the angular momentum control of the incident vector vortex beams, the spin skyrmions can also provide a new platform to study strong spin-orbital interaction~\cite{li2020mapping,shi2020strong,shi2021transverse,lei2021optical}. Plasmonic spin field has recently been also used to tailor more complex skyrmionic structures, such as the merons with fractional topological charges~\cite{dai2020plasmonic,xiong2021polaritonic}, whereby a novel quasiparticle triplet-meron structure were designed and generated (Fig.~\ref{f2}\textbf{c}), with nanometric and femtosecond space-time resolution~\cite{dai2020plasmonic}.\\[4pt]

\noindent\textbf{Unifying field- and spin-skyrmion manifestations.}
The symmetry considerations completely determine the topology of the evanescent electromagnetic field either through interference or spin-orbit coupling in the case of OAM carrying beams~\cite{lei2021photonic}. Spin-skyrmion and spin-meron lattices are formed due to a broken rotational symmetry of the field with orbital angular momentum, with 6-fold symmetry being responsible for a skyrmion lattice and 4-fold symmetry for a meron lattice in the appropriately shaped resonator (Fig.~\ref{f2}\textbf{d})~\cite{lei2021photonic,ghosh2021topological,zhang2022optical,ghosh2023spin}. In the absence of the spin-orbit coupling, in the case when the field is not carrying OAM, instead of spin-skyrmions, field-skyrmion lattices are formed, thus connecting two types of optical skyrmion manifestations. The spin-skyrmion and spin-meron textures correspond to the lowest energy of the electromagnetic field configuration, therefore, energetically stable.

Topological protection for spin-skyrmion quasiparticles is ensured by spin-orbit coupling in the evanescent field. The propagation direction of a guided mode and the direction of its transverse spin (due to the longitudinal field component) are intimately locked, and one cannot be changed without the other. This applies for both planar and structured guided wave propagation. In the evanescent field carrying OAM, this results in the topological protection of of a skyrmionic spin distribution, in the absence of catastrophic events, such as out of plane scattering in the propagating waves in the adjacent media.    
Similar topological considerations apply to the field-skyrmion lattices as the propagation of the guided modes and spins are linked with the additional requirement for the field symmetry, restricting the existence of individual field skyrmions.    
\\[4pt] 

\noindent
\textbf{Skyrmions and merons in structured media}\\[2pt]

\noindent\textbf{Skyrmions in the presence of chirality}.
As discussed above, the optical skyrmions in evanescent fields are limited to N\'eel-type topological textures, and it is a challenge to generate optical Bloch-type skyrmions at the interface of homogeneous media. In contrast to Neel-type skyrmions, the Bloch-type also requires helicity, therefore, the control of Bloch-type skyrmions can bring additional degree of freedom to control more diversified quasiparticle textures. This limitation can be overcome exploiting optical fields in complex, spatially modulated media. Bloch-type spin-skyrmions were theoretically proposed in a designer multilayer chiral liquid crystal structure~\cite{zhang2021bloch}, in which chirality-dependent twisted features occur in the spin vector field. The required vorticity can be achieved as an additional degree of freedom for extending applications to optical chiral sensing and data storage technologies. The meron textures of the polarization Stokes field was experimentally generated by confining light into a liquid-crystal-filled microcavity 
(Fig.~\ref{f2}\textbf{e})~\cite{krol2021observation}. Anisotropy of the medium may provide additional degree of freedom to develop new topological skyrmionic states in the future.\\[4pt] 

\noindent\textbf{Pseudospin skyrmions in nonlinear media.}
Besides the SAM vectors, other kinds of optical vector field in complex media can be considered to construct skyrmions. One of the implementations is to use pseudospin vectors in nonlinear photonic crystals~\cite{karnieli2021emulating}. Here, the pseudospin is defined as the wavevector involving two interacting frequencies (idler and signal frequencies) in nonlinear conversion which can be mapped onto a Bloch sphere. This results in an effective "magnetization" of an especially-designed nonlinear medium (Fig.~\ref{f2}\textbf{f})~\cite{karnieli2022geometric,karnieli2018all}.
This proposal offers the possibility to synthesize higher-order optical skyrmions.
The interaction between a light beam and the pseudospin skyrmion with controlled higher-order texture in a nonlinear crystal can be used to emulate properties of magnetic skyrmions in materials, i.e. spin transport and topologically protected skyrmion Hall effect, promising a powerful tool for applications of new photonic Hall devices~\cite{karnieli2021emulating,jiang2017direct,chen2017skyrmion}.\\[4pt] 

\noindent
\textbf{Skyrmions in free-space fields}\\[2pt]
\\
\noindent\textbf{Skyrmions in tightly focused beams.}
In paraxial beams, the electromagnetic field is a purely transverse in non-absorbing media, therefore, the field vectors always lie in a plane normal to the propagation direction. 
However, the tight focusing may introduce a longitudinal field component in the field structure. In this case, 3D vector distribution can be achieved in a propagating field, which can be used to construct skyrmionic topologies. The spin vectors in a tightly focused vortex beam may look like a N\'eel-type skyrmion-like texture in the focal plane near the beam axis, however, it is not a rigorous analogue of a skyrmion and a skyrmion number can not be assigned~\cite{du2019deep}. The reason is that a diffraction-free propagating beam is boundary-free and, therefore, is unable to form a well-defined topological structure, in contrast to an evanescent optical vortexes. 

Optical skyrmions in tightly focused fields can be designed using complex structured light, such as the full-Poincar\'e vector (FPV) beams, as the input light. Tight focusing was shown to enable optical spin skyrmions where N\'eel- and Bloch-type textures can be controlled by slightly tuning the input light (Fig.~\ref{f2}\textbf{g})~\cite{gutierrez2021optical}.
The field-skyrmion topologies in such propagating beams may theoretically exhibit strong stability against scattering in disordered  media~\cite{liu2022disorder}.\\[4pt]

\noindent\textbf{Stokes skyrmions in paraxial beams.}
 The structured vector beams with spatially variant polarization has the ability to form 3D Stokes vector distributions. A 3D Stokes vector with three components being the three Stokes parameters, describes the polarisation state, generally a 2D polarization ellipse. A Stokes skyrmion was introduced corresponding to a spatial variation of a polarization pattern in a vector beam~\cite{gao2020paraxial,gao2020e}. This can be seen as a polarization pattern unwrapped from the Poincar\'e sphere. Such kind of topological vector beams has been proposed as realisations of full Poincar\'e beams a decade ago~\cite{beckley2010full}, but only recently being related to skyrmions~\cite{donati2016twist,gao2020paraxial,shen2021topological}. To be classified as a skyrmion, the vector texture described by a topological number must be protected upon propagation. Such skyrmionic beams were identified considering superposition of a vortex beam and a non-vortex beam with orthogonal polarizations, i.e. a nonseparable state between spatial mode of polarization, where the two modal components must have a fixed ratio of the Gouy phase and aligned at the beam waist~\cite{gao2020paraxial,gao2020e,shen2021topological}. If the modes have misaligned Gouy phases, the beam will not be topologically protected and show variant topological charge upon propagation. 

Topologically protected free-space skyrmionic beams play an important role for controlling vector structure of optical beams. In particular, the digital hologram method can be used to generate tuneable skyrmions with all the textures of N\'eel-, Bloch, and anti-types~\cite{shen2021generation} (Fig.~\ref{f2}\textbf{g}). The Stokes skyrmions can also be generated directly in lasers with skyrmion numbers controlled by the resonator grating~\cite{lin2021microcavity} (Fig.~\ref{f2}\textbf{h}).  

A unique feature of Stokes-skyrmionic vector beams is the propagation-dependent topology. Upon the propagation of a skyrmionic beam, the vector texture can evolve from Bloch-type into N\'eel-type and vise versa in free space. However, in the case of an anti-skyrmion texture, the vector field texture will not change during propagation~\cite{shen2021topological}. This principle is similar to the topological protection of original skyrmions in particle physics, as Bloch-type and N\'eel-type skyrmionic textures are in the same topologically protected group, while an anti-skyrmion is in a different group.

It is worth noting that the Stokes skyrmionic beams have formal overlaps with the polarization singularities in singular optics~\cite{dennis2009singular}, but arise from entirely different considerations. Firstly, skyrmions require the continuous spin textures which consist of fully 3D oriented spin vectors in a localized region, while singular optics involves only transverse patterns of polarization ellipse and their discontinuities. Although, some transversely projected patterns of skyrmionic fields can exhibit similarity with singular polarization patterns (C-points, Lemons, Stars), they, however, originate from a very different theoretical description and are not topologically protected.

Free-space skyrmions possess topological protection in analogy to the topological protection of skyrmions in magnetic materials. Once a magnetic skyrmion is formed in a ferromagnet, its topology in a stable state is supported by the nonlinear interactions among magnetic spins and difficult to destroy by external perturbations. Similarly, once a optical skyrmionic pattern is designed in a free-space by the mode nonseparability and Gouy-phase locking discussed above, the topology and skyrmion number is invariant upon light propagation and resilient to perturbations, except for catastrophic events, such as strong scattering. This topological protection was demonstrated as a salient property of vectorial structured light due to its polarization inhomogeneity, in contrast to scalar structured light~\cite{nape2022revealing}. When such a space-polarization nonseparable vector beam propagates in complex media, the scattering can easily distort scalar mode components and local polarization, however, the global topology related to the orthogonality and nonseparablity of spatial mode and polarization is very difficult to change~\cite{nape2022revealing}. Such topological protection was recently theoretically and experimentally demonstrated for different kinds of perturbations in complex media such as liquid and atmosphere turbulence, laying the base for applying free-space skyrmions as a next-generation robust carrier for information transfer~\cite{klug2023robust,nape2022revealing}. 
\\[4pt]

\noindent
\textbf{Skyrmions in space-time}\\[2pt]
\\
Until now, we considered free-space skyrmions in continuous-wave free-space light, where the skyrmion textures are observed in a certain transverse plane. This description can also be applied to skyrmions with not too short optical pulses which are space-time separable. The situation is different in the case of the few-cycle ultrafast pulses, which represent space-time nonseparable solutions of the Maxwell's equations~\cite{zdagkas2020space}. In the latter case, propagating skyrmions can be achieved with the toroidal light pulses~\cite{hellwarth1996focused}. Such pulses induce multiple singularities in their electromagnetic field, which form vortex-ring-like configurations, and skyrmionic topology can be observed in their transverse cross-section. The experimental generation of toroidal pulses can be achieved using a radially polarised pulse interaction with a gradient metasurface designed to achieve space-time coupling~\cite{zdagkas2021observation}. Interestingly, there exist also another kind of optical toroidal pulses that belong to a scalar light field with the phase gradient lines resembling skyrmion structure rather than the real optical vector field~\cite{wan2022toroidal,wan2022scalar}.The toroidal pulses can be generalized into supertoroidal, fractal-like configurations of singularity distributions and vortex rings~\cite{shen2021supertoroidal}. With the increasing supertoroidal order parameter, multiple propagating skyrmions can be obtained with different topological textures (polarity and helicity) within the pulse~\cite{shen2021supertoroidal} (Fig.~\ref{f2}\textbf{i}). 
The related skyrmions in the pulse always exhibit a constant vorticity of 1, while the polarity alternates between $+1$ and $-1$ due to the optical cycle evolution. Diffractionless propagation of supertoroidal pulses may potentially enable the skyrmion topologies to be transported over an arbitrary long distance~\cite{shen2022nondiffracting}.
The higher-order skyrmions in ultrashort pulses are still to be explored.
\\[4pt]

\noindent
\textbf{Skyrmions in momentum-space}\\[2pt]
\\
\noindent
Skyrmions in momentum ($k$) space are defined in a similar way as those in real space, only replacing the real space for
the wavevector space for topological mapping. 
Spin textures in a momentum space provide a direct approach to visualizing the local Berry curvature, enabling robust and direct manipulation of polarization of light~\cite{guo2020meron} and space-time coupling of light wave packets~\cite{guo2021structured}. The momentum-space magnetic skyrmions were previously studied to demonstrate anomalous Hall effect in magnets~\cite{nagaosa2012gauge,loder2017momentum}.

  Momentum-space spin textures in photonic systems promise intriguing physical insights related to Dirac monopoles and photon spin quantization, whereby the magnetic monopole charge quantization in momentum space can arise from spin-1 skyrmion properties of polaritons.~\cite{van2019photonic}. The first implementation of optical quasiparticles in momentum-space was demonstrated by engineering the band structure of a photonic crystal with a broken symmetry to enable meron textures formed by Stokes vectors in the Brillouin zone (Fig.~\ref{f2}\textbf{j})~\cite{guo2020meron}.
Such momentum-space meron was used for generating space-time light bullets, i.e. the 3D localized wave packets with complex space-time coupling and nondiffracting propagation. In particular, a circularly polarized Gaussian pulse was injected into a photonic slab whose band structure was engineered with a meron texture in its $k$-space, and then can be converted as a light bullet with a complex spatiotemporal polarization structure which possess a Stokes meron texture in its wave-vector space.~\cite{guo2021structured}. 
\\[4pt]

\noindent
\textbf{Optical quasiparticles beyond skyrmions and merons}\\
In the previous sections, we discussed 2D skyrmions and merons described by a fundamental topological numbers of polarity, vorticity, and helicity. More complex cases of topological quasiparticles are also possible, such as a skyrmionium, which requires a radial twisting number, a bimeron obtained with a tilted parameter sphere and unwrapping from the equator, as well as 3D quasiparticles (Box~1).

\noindent\textbf{Optical skyrmionium and target skyrmions.}
In addition to fundamental order skyrmions, other generalized quasiparticles of electromagnetic field topology have recently been demonstrated. 
Typical examples are skyrmioniums and $k\pi$ target skyrmions with radial multi-twist structures. 
The $k\pi$ skyrmions were reported in free-space skyrmionic vector beams with reconfigurable topological textures, including higher-order vorticity and $k\pi$ radial twist~\cite{shen2021generation}. More complex optical quasiparticles with arbitrary hybrid higher-order $k\pi$ target skyrmions may be expected.\\[4pt]
\\
\noindent\textbf{Optical bimerons, bimeroniums, and multiskyrmions.}
Bimerons can be seen as a class of generalized quasiparticle isomorphic to skyrmions, which were reported in solid-state physics and chiral magnets. While the optical bimerons were only recently first proposed in paraxial vector beams~\cite{shen2021topological}. The bimeronic beam can be obtained by modifying a skyrmionic beam, which requires a superposition with two specially designed spatial modes with two different non-orthogonal polarizations (a skyrmionic beam requires them to be orthogonal). In this case, the degree of the non-orthogonality controls the tilt of the parameter sphere to obtain a bimeron (Box~1). 

Similarly, bimeronium is a quasiparticle which can be obtained by adding radial $k\pi$ twist structure to a fundamental bimeron (akin to the concept of skyrmionium)~\cite{zhang2021frustrated}. An optical bimeronium can be realized by replacing the scalar mode bases of a bimeronic beam with higher-order vortex modes or Bessel beams. The Stokes bimerons were recently generalized into multiskyrmion structures with increasingly complex topologies~\cite{cisowski2023building,shen2023topologically}.\\[4pt]


\noindent\textbf{3D skyrmion tube, hopfion, and beyond.}
While primarily 2D skyrmionic structures were widely considered in the electromagnetic domain, the studies of 3D skyrmions have also been recently initiated in the structured light. A 3D analysis of the Stokes vector field of a paraxial full-Poincar\'e vector beam  reveals an analogue of a skyrmion tube~\cite{kuratsuji2021evolution}. Using a hypersphere as a higher-dimensional generalization of a Poincar\'e sphere, the unwrapping of the vector field structure results in a skyrmionic texture in 3D which was termed an optical hopfion~\cite{sugic2021particle}. In such optical hopfion, each fibre in the Hopf fibration corresponds to the trajectory of a certain polarization ellipse~\cite{sugic2021particle} (Fig.~\ref{f2}\textbf{k}). The optical hopfions were recently extended to higher-order quasiparticles and their topologically protected transport in free space was demonstrated~\cite{shen2023topological}. There are other optical counterparts of 3D quasiparticles to be explored including chiral bobbers~\cite{zheng2018experimental}, skyrmion bundles~\cite{tang2021magnetic}, heliknotons~\cite{tai2019three} and torons~\cite{ackerman2017diversity}.
\\[4pt]


\noindent
\textbf{\large Potential applications}\\[2pt]
\noindent
\\
The important features of optical skyrmionic textures can be summarized as:
(i) ultrasmall- the spatial size of a spin-skyrmion features  can be realized on a  deep-subwavelength scale as the spin is not subject to diffraction ~\cite{du2019deep,davis2020ultrafast,dai2020plasmonic}; (ii) ultrafast- the skyrmion texture can be resolved on femtosecond scales and at optical single-cycle field level~\cite{dai2020plasmonic,shen2021supertoroidal}; (iii) diversity of topologies- the vector textures of skyrmions can be characterized by multiple topological numbers and extended to higher dimensional cases~\cite{shen2021topological,shen2021generation}; (iv) topological protection- the topology of a photonic skyrmion texture can be very stable and resilient to disorder and perturbation in complex media unless catastrophic events~\cite{liu2022disorder,nape2022revealing}. The ultrasmall and ultrafast features play an important role in developing applications in super-resolution imaging and metrology beyond the diffraction limit. The diversity and stability of particle-like topologies allows consideration of the particle physics and condensed matter physics effects in optics in order to study and tailor light-matter interactions for quantum and topological technologies.\\[4pt]

\noindent\textbf{Light-matter interactions}
\\
In condensed matter physics, skyrmions can be formed due to the Dzyaloshinskii–Moriya interaction (DMI)~\cite{yang2023first}, dipole–dipole interaction~\cite{zhang2020skyrmion} and can be driven by the skyrmion Hall effect~\cite{jiang2017direct,chen2017skyrmion}, spin-orbit torque~\cite{vzelezny2018spin,yang2021chiral}, and even the light~\cite{sohn2019light,sohn2020optically,fujita2017ultrafast,yang2018photonic,hirosawa2022laser}. The exact Hamiltonian for the interactions leading to photonic skyrmionic structures is yet to be fully formulated. This would be an exciting development which will allow not only to study the optical counterparts of solid-state skyrmionic phenomena but predict new ones specific to optics. Recently, an electric analogue of the DMI was demonstrated~\cite{zhao2021dzyaloshinskii,junquera2021dzyaloshinskii}, which inspires exploration of the optical DMI.  

Similar mechanisms can be studied in skyrmionic electromagnetic light fields. For isolated spin-skyrmions, optical analog of the exchange energy and the DMI terms are related to conservation of the total angular momentum and the spin texture stability is protected by the optical spin-Hall effect and symmetry of the optical system in the presence of OAM~\cite{shi}. Skyrmion Hall effect recently found its optical counterpart in pseudospin skyrmions in nonlinear media, which lays foundation to design novel topological photonic devices~\cite{karnieli2021emulating}. Combining the nonlinear processes with spin-to-orbital coupling in the structured light, skyrmionic evolution in optical nonlinear frequency conversion is also an important emerging topic~\cite{wu2022conformal,watzel2020topological}. In turn, applying optical concepts to magnetism, magnetic skyrmions have been used as a magnetic tweezers with a similar mechanism to optical tweezers with twisted light~\cite{jiang2020twisted,wang2020optical,padgett2011tweezers}. Optical skyrmions can interact with magnetic skyrmions allowing for a route to manipulate these magnetic particles~\cite{fujita2017ultrafast,yang2018photonic,hirosawa2022laser}. Recently, it has also been experimentally demonstrated that the interactions between light and topological quasiparticles in liquid crystals can enable local transfer of momentum between light and matter, revealing a new mechanism to develop advanced applications such as solitonic tractors and nonlinear photonic devices~\cite{poy2022interaction}. \\[4pt]


\noindent\textbf{Microscopy, metrology, and sensing}
\\
Optical plasmonic field- and spin-skyrmions were verified to possess deep-subwavelength features with sufficient light intensities~\cite{du2019deep,davis2020ultrafast}, which meets the emerging requirements of superresolution imaging and metrology. In contrast to intensity variations, the local spin distribution in the skyrmions varies at the deep-subwavelength scales, down to $\lambda/60$ level or $\sim10$~nm~\cite{du2019deep}. Much higher resolution is possible at the expense of the reduced light intensity required for detection. The fine structure of spin skyrmion can be used as a tool for imaging, promising a superresolution microscopic method beyond diffraction limit. This would be especially advantageous for imaging the material properties influencing optical polarisation, such as magnetic or ferroelectric domain distributions~\cite{lei2021optical}. This is turn promises ultraprecise detection of magnetic skyrmions~\cite{tengdin2021imaging}.
The plasmonic field skyrmions were also applied for frequency sensing, achieving high sensitivity ($>4.82\%$) and extreme accuracy ($\approx99.99\%$) in the deep-subwavelength areas 26$\times$ 10$^{-5}\lambda^2$~\cite{li2022highly}, which is a milestone for ultracompact highly sensitive sensors. The optical spin skyrmions were also recently used in developing ultraprecise displacement sensing or metrology achieving pico-metric resolution~\cite{yang2023spin}.\\[4pt]

The resolving of fine structure of light and control of its movement also serves as basic scale mark to enable ultraprecise displacement metrology, such as the optical ruler metrology~\cite{yuan2019detecting}. By combining a pair of $\pm1$st-order spin skyrmions with an accurately controlled separation, a structured spin field akin to the field of an electron-positron pair can be constructed and a displacement sensing technology with sensitivity down to angstrom/picometer and effective linear range exceeding hundreds of nanometer was successfully realized and the localization can be achieved by measuring the spin degree of such a structured spin distribution~\cite{du2021ultrasensitive}. 
\\[4pt]

\noindent\textbf{Optical communications and informatics}
\\
The most important applications of magnetic skyrmions are the high-density information storage in modern spintronics~\cite{zhang2020skyrmion}. Topological parameters of skyrmions have great advantage to encode and process information. Similarly, in optics, the development of optical information storage and communication may be enabled by structured light, in particular using vector vortex beams~\cite{shen2019optical}. Optical skyrmions stand at the crossroads of spintronics and structured light and may lead to new protocols in optical communications and informatics, by transferring methods from spintronics to structured light and utilising various degrees of freedom provided by topologically protected vectorial field textures. In magnetic skyrmionic racetrack memory for high-density information storage, the information is encoded in the presence and absence of skyrmions~\cite{zhang2020skyrmion}. 
Similar topology-based protocols of information encoding, processing and transport can be envisaged in optics.

For quantum technologies, since skyrmions are characterised by multiple topological charges, e.g. polarity and vorticity, exotic biparticle entangled states can be considered with more degrees of freedom. For instance, a skyrmion of polarity-up and vorticity of $\ell$ can be entangled with another skyrmion of polarity-up and vorticity of $-\ell$. More complex skyrmion textures can be involved in the entanglement, such as the skyrmionium or $k\pi$ skyrmion described by additional topological numbers, to access hybrid-entangled states with multiple degrees of freedom, beyond spin or OAM entangled photons. A nonlocal optical skyrmion as a quantum entangled state of two photons with
hybrid degrees of freedom, spatial mode and polarization, fulfilling a skyrmion mapping in the correlation function of the two
entangled photons has recently been introduced~\cite{ornelas2022non} (Fig.~\ref{f2}\textbf{l}).
\\[4pt]

\noindent
\textbf{\large Outlook}\\[2pt]
\noindent
\\
The realisation of optical counterparts of skyrmion quasiparticles has enabled the novel topological approach for controlling electromagnetic fields. Despite the continuous stream of elegant experimental and theoretical results, the field is still in its infancy with many fundamental theoretical questions to be resolved and applications to be explored. 

Variety of optical field characteristics have been already used to construct topological optical quasiparticles, but there is still a broad parameter space to explore. There are still plenty of sophisticated topological textures predicted in particle physics to be studied in optical fields.
With the rapid development of structured light technologies and optical field shaping in all its degrees of freedom and dimensions, the combination of the field topologies and matter topologies will be on a table. They include both natural topological states of matter, such as magnetic skyrmions, as well as artificially structured materials with engineered topologies. The interaction of the ultrafast space-time topological states with time-varying media~\cite{galiffi2022photonics}, which provide artificial gauge fields, is a new frontier of topological science.
Optical quasiparticles considered in this review in a classical optics, while, the study of quantum and nonlocal skyrmionic states is also emerging. The complex topology of skyrmions may provide great potential to explore higher-dimensional quantum states for quantum technologies which are difficult to reach otherwise.
The topological approaches developed for optical skyrmions can also be transferred to other wave fields such as acoustic (sound) waves~\cite{ge2021observation,muelas2022observation,hu2023observation}, elastic waves~\cite{cao2023observation}, and atomic waves~\cite{parmee2022optical}. Always, however, care should be taken to distinguish vector textures exhibiting true topological properties of quasiparticles from their lookalikes without topological protection. Topological considerations of waves become increasingly crucial for meeting the requirements of ultrabroad bandwidth information transfer, high-resolution optical microscopy and metrology, precise control of optical forces, magnetic and nonlinear processes, as well as quantum technologies. \\[4pt]

\bibliographystyle{naturemag}

\bigskip
\noindent\textbf{\large Acknowledgments}\\
The authors would like to thank Cheng Guo for useful discussions and Pedro Ornelas for assisting with graphics. This work is funded by National Natural Science Foundation of China (grants No. 12047540, U1701661, 61935013, 62075139, and 12174266), Guangdong Major Project of Basic Research No. 2020B0301030009, and European Research Council iCOMM project (789340).

\bigskip
\noindent\textbf{\large Competing interests}\\
The authors declare no competing interests.

\bibliography{sample}

\providecommand{\noopsort}[1]{}\providecommand{\singleletter}[1]{#1}%
\begin{thebibliography}{100}
\expandafter\ifx\csname url\endcsname\relax
  \def\url#1{\texttt{#1}}\fi
\expandafter\ifx\csname urlprefix\endcsname\relax\def\urlprefix{URL }\fi
\providecommand{\bibinfo}[2]{#2}
\providecommand{\eprint}[2][]{\url{#2}}

\bibitem{skyrme1991non}
\bibinfo{author}{Skyrme, T. H.~R.}
\newblock \bibinfo{title}{A non-linear field theory}.
\newblock \emph{\bibinfo{journal}{Proc. R. Soc. A}}
  \textbf{\bibinfo{volume}{260}}, \bibinfo{pages}{127--138}
  (\bibinfo{year}{1961}).

\bibitem{skyrme1962unified}
\bibinfo{author}{Skyrme, T. H.~R.}
\newblock \bibinfo{title}{A unified field theory of mesons and baryons}.
\newblock \emph{\bibinfo{journal}{Nuclear Physics}}
  \textbf{\bibinfo{volume}{31}}, \bibinfo{pages}{556--569}
  (\bibinfo{year}{1962}).

\bibitem{zahed1986skyrme}
\bibinfo{author}{Zahed, I.} \& \bibinfo{author}{Brown, G.~E.}
\newblock \bibinfo{title}{The skyrme model}.
\newblock \emph{\bibinfo{journal}{Physics Reports}}
  \textbf{\bibinfo{volume}{142}}, \bibinfo{pages}{1--102}
  (\bibinfo{year}{1986}).

\bibitem{naya2018skyrmions}
\bibinfo{author}{Naya, C.} \& \bibinfo{author}{Sutcliffe, P.}
\newblock \bibinfo{title}{Skyrmions and clustering in light nuclei}.
\newblock \emph{\bibinfo{journal}{Physical Review Letters}}
  \textbf{\bibinfo{volume}{121}}, \bibinfo{pages}{232002}
  (\bibinfo{year}{2018}).

\bibitem{halcrow2020attractive}
\bibinfo{author}{Halcrow, C.} \& \bibinfo{author}{Harland, D.}
\newblock \bibinfo{title}{Attractive spin-orbit potential from the skyrme
  model}.
\newblock \emph{\bibinfo{journal}{Physical Review Letters}}
  \textbf{\bibinfo{volume}{125}}, \bibinfo{pages}{042501}
  (\bibinfo{year}{2020}).

\bibitem{al2001skyrmions}
\bibinfo{author}{Al~Khawaja, U.} \& \bibinfo{author}{Stoof, H.}
\newblock \bibinfo{title}{Skyrmions in a ferromagnetic bose--einstein
  condensate}.
\newblock \emph{\bibinfo{journal}{Nature}} \textbf{\bibinfo{volume}{411}},
  \bibinfo{pages}{918--920} (\bibinfo{year}{2001}).

\bibitem{duzgun2021skyrmion}
\bibinfo{author}{Duzgun, A.} \& \bibinfo{author}{Nisoli, C.}
\newblock \bibinfo{title}{Skyrmion spin ice in liquid crystals}.
\newblock \emph{\bibinfo{journal}{Physical Review Letters}}
  \textbf{\bibinfo{volume}{126}}, \bibinfo{pages}{047801}
  (\bibinfo{year}{2021}).

\bibitem{liu2016skyrmions}
\bibinfo{author}{Liu, J.~P.}, \bibinfo{author}{Zhang, Z.} \&
  \bibinfo{author}{Zhao, G.}
\newblock \emph{\bibinfo{title}{Skyrmions: topological structures, properties,
  and applications}} (\bibinfo{publisher}{CRC Press}, \bibinfo{year}{2016}).

\bibitem{tokura2020magnetic}
\bibinfo{author}{Tokura, Y.} \& \bibinfo{author}{Kanazawa, N.}
\newblock \bibinfo{title}{Magnetic skyrmion materials}.
\newblock \emph{\bibinfo{journal}{Chemical Reviews}}
  \textbf{\bibinfo{volume}{121}}, \bibinfo{pages}{2857--2897}
  (\bibinfo{year}{2021}).

\bibitem{bogdanov2020physical}
\bibinfo{author}{Bogdanov, A.~N.} \& \bibinfo{author}{Panagopoulos, C.}
\newblock \bibinfo{title}{Physical foundations and basic properties of magnetic
  skyrmions}.
\newblock \emph{\bibinfo{journal}{Nature Reviews Physics}}
  \textbf{\bibinfo{volume}{2}}, \bibinfo{pages}{492--498}
  (\bibinfo{year}{2020}).

\bibitem{fert2017magnetic}
\bibinfo{author}{Fert, A.}, \bibinfo{author}{Reyren, N.} \&
  \bibinfo{author}{Cros, V.}
\newblock \bibinfo{title}{Magnetic skyrmions: advances in physics and potential
  applications}.
\newblock \emph{\bibinfo{journal}{Nature Reviews Materials}}
  \textbf{\bibinfo{volume}{2}}, \bibinfo{pages}{1--15} (\bibinfo{year}{2017}).

\bibitem{lima2022spin}
\bibinfo{author}{Lima~Fernandes, I.}, \bibinfo{author}{Bl{\"u}gel, S.} \&
  \bibinfo{author}{Lounis, S.}
\newblock \bibinfo{title}{Spin-orbit enabled all-electrical readout of chiral
  spin-textures}.
\newblock \emph{\bibinfo{journal}{Nature Communications}}
  \textbf{\bibinfo{volume}{13}}, \bibinfo{pages}{1--10} (\bibinfo{year}{2022}).

\bibitem{han2022high}
\bibinfo{author}{Han, L.} \emph{et~al.}
\newblock \bibinfo{title}{High-density switchable skyrmion-like polar
  nanodomains integrated on silicon}.
\newblock \emph{\bibinfo{journal}{Nature}} \textbf{\bibinfo{volume}{603}},
  \bibinfo{pages}{63--67} (\bibinfo{year}{2022}).

\bibitem{khalaf2021charged}
\bibinfo{author}{Khalaf, E.}, \bibinfo{author}{Chatterjee, S.},
  \bibinfo{author}{Bultinck, N.}, \bibinfo{author}{Zaletel, M.~P.} \&
  \bibinfo{author}{Vishwanath, A.}
\newblock \bibinfo{title}{Charged skyrmions and topological origin of
  superconductivity in magic-angle graphene}.
\newblock \emph{\bibinfo{journal}{Science Advances}}
  \textbf{\bibinfo{volume}{7}}, \bibinfo{pages}{eabf5299}
  (\bibinfo{year}{2021}).

\bibitem{kwan2022skyrmions}
\bibinfo{author}{Kwan, Y.~H.}, \bibinfo{author}{Wagner, G.},
  \bibinfo{author}{Bultinck, N.}, \bibinfo{author}{Simon, S.~H.} \&
  \bibinfo{author}{Parameswaran, S.}
\newblock \bibinfo{title}{Skyrmions in twisted bilayer graphene: stability,
  pairing, and crystallization}.
\newblock \emph{\bibinfo{journal}{Physical Review X}}
  \textbf{\bibinfo{volume}{12}}, \bibinfo{pages}{031020}
  (\bibinfo{year}{2022}).

\bibitem{muhlbauer2009skyrmion}
\bibinfo{author}{M{\"u}hlbauer, S.} \emph{et~al.}
\newblock \bibinfo{title}{Skyrmion lattice in a chiral magnet}.
\newblock \emph{\bibinfo{journal}{Science}} \textbf{\bibinfo{volume}{323}},
  \bibinfo{pages}{915--919} (\bibinfo{year}{2009}).

\bibitem{yu2010real}
\bibinfo{author}{Yu, X.} \emph{et~al.}
\newblock \bibinfo{title}{Real-space observation of a two-dimensional skyrmion
  crystal}.
\newblock \emph{\bibinfo{journal}{Nature}} \textbf{\bibinfo{volume}{465}},
  \bibinfo{pages}{901--904} (\bibinfo{year}{2010}).

\bibitem{tsesses2018optical}
\bibinfo{author}{Tsesses, S.} \emph{et~al.}
\newblock \bibinfo{title}{Optical skyrmion lattice in evanescent
  electromagnetic fields}.
\newblock \emph{\bibinfo{journal}{Science}} \textbf{\bibinfo{volume}{361}},
  \bibinfo{pages}{993--996} (\bibinfo{year}{2018}).

\bibitem{du2019deep}
\bibinfo{author}{Du, L.}, \bibinfo{author}{Yang, A.}, \bibinfo{author}{Zayats,
  A.~V.} \& \bibinfo{author}{Yuan, X.}
\newblock \bibinfo{title}{Deep-subwavelength features of photonic skyrmions in
  a confined electromagnetic field with orbital angular momentum}.
\newblock \emph{\bibinfo{journal}{Nature Physics}}
  \textbf{\bibinfo{volume}{15}}, \bibinfo{pages}{650--654}
  (\bibinfo{year}{2019}).

\bibitem{rivera2020light}
\bibinfo{author}{Rivera, N.} \& \bibinfo{author}{Kaminer, I.}
\newblock \bibinfo{title}{Light--matter interactions with photonic
  quasiparticles}.
\newblock \emph{\bibinfo{journal}{Nature Reviews Physics}}
  \textbf{\bibinfo{volume}{2}}, \bibinfo{pages}{538--561}
  (\bibinfo{year}{2020}).

\bibitem{manton2004topological}
\bibinfo{author}{Manton, N.} \& \bibinfo{author}{Sutcliffe, P.}
\newblock \emph{\bibinfo{title}{Topological solitons}}
  (\bibinfo{publisher}{Cambridge University Press}, \bibinfo{year}{2004}).

\bibitem{hopf1964abbildungen}
\bibinfo{author}{Hopf, H.}
\newblock \bibinfo{title}{{\"U}ber die abbildungen der dreidimensionalen
  sph{\"a}re auf die kugelfl{\"a}che}.
\newblock \emph{\bibinfo{journal}{Mathematische Analen}}
  \textbf{\bibinfo{volume}{104}}, \bibinfo{pages}{637--665}
  (\bibinfo{year}{1931}).

\bibitem{faddeev1976some}
\bibinfo{author}{Faddeev, L.}
\newblock \bibinfo{title}{Some comments on the many-dimensional solitons}.
\newblock \emph{\bibinfo{journal}{Letters in Mathematical Physics}}
  \textbf{\bibinfo{volume}{1}}, \bibinfo{pages}{289--293}
  (\bibinfo{year}{1976}).

\bibitem{gobel2021beyond}
\bibinfo{author}{G{\"o}bel, B.}, \bibinfo{author}{Mertig, I.} \&
  \bibinfo{author}{Tretiakov, O.~A.}
\newblock \bibinfo{title}{Beyond skyrmions: Review and perspectives of
  alternative magnetic quasiparticles}.
\newblock \emph{\bibinfo{journal}{Physics Reports}}
  \textbf{\bibinfo{volume}{895}}, \bibinfo{pages}{1--28}
  (\bibinfo{year}{2021}).

\bibitem{nagaosa2013topological}
\bibinfo{author}{Nagaosa, N.} \& \bibinfo{author}{Tokura, Y.}
\newblock \bibinfo{title}{Topological properties and dynamics of magnetic
  skyrmions}.
\newblock \emph{\bibinfo{journal}{Nature Nanotechnology}}
  \textbf{\bibinfo{volume}{8}}, \bibinfo{pages}{899--911}
  (\bibinfo{year}{2013}).

\bibitem{kezsmarki2015neel}
\bibinfo{author}{K{\'e}zsm{\'a}rki, I.} \emph{et~al.}
\newblock \bibinfo{title}{N{\'e}el-type skyrmion lattice with confined
  orientation in the polar magnetic semiconductor gav 4 s 8}.
\newblock \emph{\bibinfo{journal}{Nature Materials}}
  \textbf{\bibinfo{volume}{14}}, \bibinfo{pages}{1116--1122}
  (\bibinfo{year}{2015}).

\bibitem{milde2013unwinding}
\bibinfo{author}{Milde, P.} \emph{et~al.}
\newblock \bibinfo{title}{Unwinding of a skyrmion lattice by magnetic
  monopoles}.
\newblock \emph{\bibinfo{journal}{Science}} \textbf{\bibinfo{volume}{340}},
  \bibinfo{pages}{1076--1080} (\bibinfo{year}{2013}).

\bibitem{nayak2017magnetic}
\bibinfo{author}{Nayak, A.~K.} \emph{et~al.}
\newblock \bibinfo{title}{Magnetic antiskyrmions above room temperature in
  tetragonal heusler materials}.
\newblock \emph{\bibinfo{journal}{Nature}} \textbf{\bibinfo{volume}{548}},
  \bibinfo{pages}{561--566} (\bibinfo{year}{2017}).

\bibitem{zhang2017direct}
\bibinfo{author}{Zhang, S.}, \bibinfo{author}{Van Der~Laan, G.} \&
  \bibinfo{author}{Hesjedal, T.}
\newblock \bibinfo{title}{Direct experimental determination of the topological
  winding number of skyrmions in cu 2 oseo 3}.
\newblock \emph{\bibinfo{journal}{Nature Communications}}
  \textbf{\bibinfo{volume}{8}}, \bibinfo{pages}{1--7} (\bibinfo{year}{2017}).

\bibitem{zhang2016control}
\bibinfo{author}{Zhang, X.} \emph{et~al.}
\newblock \bibinfo{title}{Control and manipulation of a magnetic skyrmionium in
  nanostructures}.
\newblock \emph{\bibinfo{journal}{Physical Review B}}
  \textbf{\bibinfo{volume}{94}}, \bibinfo{pages}{094420}
  (\bibinfo{year}{2016}).

\bibitem{kolesnikov2018skyrmionium}
\bibinfo{author}{Kolesnikov, A.~G.}, \bibinfo{author}{Stebliy, M.~E.},
  \bibinfo{author}{Samardak, A.~S.} \& \bibinfo{author}{Ognev, A.~V.}
\newblock \bibinfo{title}{Skyrmionium--high velocity without the skyrmion hall
  effect}.
\newblock \emph{\bibinfo{journal}{Scientific reports}}
  \textbf{\bibinfo{volume}{8}}, \bibinfo{pages}{1--8} (\bibinfo{year}{2018}).

\bibitem{song2019field}
\bibinfo{author}{Song, C.} \emph{et~al.}
\newblock \bibinfo{title}{Field-tuned spin excitation spectrum of k$\pi$
  skyrmion}.
\newblock \emph{\bibinfo{journal}{New Journal of Physics}}
  \textbf{\bibinfo{volume}{21}}, \bibinfo{pages}{083006}
  (\bibinfo{year}{2019}).

\bibitem{zheng2017direct}
\bibinfo{author}{Zheng, F.} \emph{et~al.}
\newblock \bibinfo{title}{Direct imaging of a zero-field target skyrmion and
  its polarity switch in a chiral magnetic nanodisk}.
\newblock \emph{\bibinfo{journal}{Physical Review Letters}}
  \textbf{\bibinfo{volume}{119}}, \bibinfo{pages}{197205}
  (\bibinfo{year}{2017}).

\bibitem{jani2021antiferromagnetic}
\bibinfo{author}{Jani, H.} \emph{et~al.}
\newblock \bibinfo{title}{Antiferromagnetic half-skyrmions and bimerons at room
  temperature}.
\newblock \emph{\bibinfo{journal}{Nature}} \textbf{\bibinfo{volume}{590}},
  \bibinfo{pages}{74--79} (\bibinfo{year}{2021}).

\bibitem{desplat2019paths}
\bibinfo{author}{Desplat, L.}, \bibinfo{author}{Kim, J.-V.} \&
  \bibinfo{author}{Stamps, R.}
\newblock \bibinfo{title}{Paths to annihilation of first-and second-order
  (anti) skyrmions via (anti) meron nucleation on the frustrated square
  lattice}.
\newblock \emph{\bibinfo{journal}{Physical Review B}}
  \textbf{\bibinfo{volume}{99}}, \bibinfo{pages}{174409}
  (\bibinfo{year}{2019}).

\bibitem{bera2019theory}
\bibinfo{author}{Bera, S.} \& \bibinfo{author}{Mandal, S.~S.}
\newblock \bibinfo{title}{Theory of the skyrmion, meron, antiskyrmion, and
  antimeron in chiral magnets}.
\newblock \emph{\bibinfo{journal}{Physical Review Research}}
  \textbf{\bibinfo{volume}{1}}, \bibinfo{pages}{033109} (\bibinfo{year}{2019}).

\bibitem{yu2018transformation}
\bibinfo{author}{Yu, X.} \emph{et~al.}
\newblock \bibinfo{title}{Transformation between meron and skyrmion topological
  spin textures in a chiral magnet}.
\newblock \emph{\bibinfo{journal}{Nature}} \textbf{\bibinfo{volume}{564}},
  \bibinfo{pages}{95--98} (\bibinfo{year}{2018}).

\bibitem{zhang2021frustrated}
\bibinfo{author}{Zhang, X.} \emph{et~al.}
\newblock \bibinfo{title}{A frustrated bimeronium: Static structure and
  dynamics}.
\newblock \emph{\bibinfo{journal}{Applied Physics Letters}}
  \textbf{\bibinfo{volume}{118}}, \bibinfo{pages}{052411}
  (\bibinfo{year}{2021}).

\bibitem{forbes2021structured}
\bibinfo{author}{Forbes, A.}, \bibinfo{author}{Oliveira, M.} \&
  \bibinfo{author}{Dennis, M.}
\newblock \bibinfo{title}{Structured light}.
\newblock \emph{\bibinfo{journal}{Nature Photonics}}
  \textbf{\bibinfo{volume}{15}}, \bibinfo{pages}{253--262}
  (\bibinfo{year}{2021}).

\bibitem{davis2020ultrafast}
\bibinfo{author}{Davis, T.~J.} \emph{et~al.}
\newblock \bibinfo{title}{Ultrafast vector imaging of plasmonic skyrmion
  dynamics with deep subwavelength resolution}.
\newblock \emph{\bibinfo{journal}{Science}} \textbf{\bibinfo{volume}{368}},
  \bibinfo{pages}{386} (\bibinfo{year}{2020}).

\bibitem{deng2021observation}
\bibinfo{author}{Deng, Z.-L.}, \bibinfo{author}{Shi, T.},
  \bibinfo{author}{Krasnok, A.}, \bibinfo{author}{Li, X.} \&
  \bibinfo{author}{Al{\`u}, A.}
\newblock \bibinfo{title}{Observation of localized magnetic plasmon skyrmions}.
\newblock \emph{\bibinfo{journal}{Nature Communications}}
  \textbf{\bibinfo{volume}{13}}, \bibinfo{pages}{1--7} (\bibinfo{year}{2022}).

\bibitem{yang2021symmetry}
\bibinfo{author}{Yang, J.} \emph{et~al.}
\newblock \bibinfo{title}{Symmetry-protected spoof localized surface plasmonic
  skyrmion}.
\newblock \emph{\bibinfo{journal}{Laser \& Photonics Reviews}}
  \textbf{\bibinfo{volume}{n/a}}, \bibinfo{pages}{2200007}
  (\bibinfo{year}{2022}).

\bibitem{bai2020dynamic}
\bibinfo{author}{Bai, C.}, \bibinfo{author}{Chen, J.}, \bibinfo{author}{Zhang,
  Y.}, \bibinfo{author}{Zhang, D.} \& \bibinfo{author}{Zhan, Q.}
\newblock \bibinfo{title}{Dynamic tailoring of an optical skyrmion lattice in
  surface plasmon polaritons}.
\newblock \emph{\bibinfo{journal}{Optics Express}}
  \textbf{\bibinfo{volume}{28}}, \bibinfo{pages}{10320--10328}
  (\bibinfo{year}{2020}).

\bibitem{meiler2020dynamic}
\bibinfo{author}{Meiler, T.}, \bibinfo{author}{Frank, B.} \&
  \bibinfo{author}{Giessen, H.}
\newblock \bibinfo{title}{Dynamic tailoring of an optical skyrmion lattice in
  surface plasmon polaritons: comment}.
\newblock \emph{\bibinfo{journal}{Optics Express}}
  \textbf{\bibinfo{volume}{28}}, \bibinfo{pages}{33614--33615}
  (\bibinfo{year}{2020}).

\bibitem{bai2020dynamicr}
\bibinfo{author}{Bai, C.}, \bibinfo{author}{Chen, J.}, \bibinfo{author}{Zhang,
  D.} \& \bibinfo{author}{Zhan, Q.}
\newblock \bibinfo{title}{Dynamic tailoring of an optical skyrmion lattice in
  surface plasmon polaritons: reply}.
\newblock \emph{\bibinfo{journal}{Optics Express}}
  \textbf{\bibinfo{volume}{28}}, \bibinfo{pages}{33616--33618}
  (\bibinfo{year}{2020}).

\bibitem{lei2021photonic}
\bibinfo{author}{Lei, X.} \emph{et~al.}
\newblock \bibinfo{title}{Photonic spin lattices: symmetry constraints for
  skyrmion and meron topologies}.
\newblock \emph{\bibinfo{journal}{Physical Review Letters}}
  \textbf{\bibinfo{volume}{127}}, \bibinfo{pages}{237403}
  (\bibinfo{year}{2021}).

\bibitem{li2020mapping}
\bibinfo{author}{Li, C.}, \bibinfo{author}{Shi, P.}, \bibinfo{author}{Du, L.}
  \& \bibinfo{author}{Yuan, X.}
\newblock \bibinfo{title}{Mapping the near-field spin angular momenta in the
  structured surface plasmon polariton field}.
\newblock \emph{\bibinfo{journal}{Nanoscale}} \textbf{\bibinfo{volume}{12}},
  \bibinfo{pages}{13674--13679} (\bibinfo{year}{2020}).

\bibitem{shi2020strong}
\bibinfo{author}{Shi, P.}, \bibinfo{author}{Du, L.} \& \bibinfo{author}{Yuan,
  X.}
\newblock \bibinfo{title}{Strong spin--orbit interaction of photonic skyrmions
  at the general optical interface}.
\newblock \emph{\bibinfo{journal}{Nanophotonics}} \textbf{\bibinfo{volume}{9}},
  \bibinfo{pages}{4619--4628} (\bibinfo{year}{2020}).

\bibitem{shi2021transverse}
\bibinfo{author}{Shi, P.}, \bibinfo{author}{Du, L.}, \bibinfo{author}{Li, C.},
  \bibinfo{author}{Zayats, A.~V.} \& \bibinfo{author}{Yuan, X.}
\newblock \bibinfo{title}{Transverse spin dynamics in structured
  electromagnetic guided waves}.
\newblock \emph{\bibinfo{journal}{Proceedings of the National Academy of
  Sciences}} \textbf{\bibinfo{volume}{118}} (\bibinfo{year}{2021}).

\bibitem{lei2021optical}
\bibinfo{author}{Lei, X.}, \bibinfo{author}{Du, L.}, \bibinfo{author}{Yuan, X.}
  \& \bibinfo{author}{Zayats, A.~V.}
\newblock \bibinfo{title}{Optical spin--orbit coupling in the presence of
  magnetization: photonic skyrmion interaction with magnetic domains}.
\newblock \emph{\bibinfo{journal}{Nanophotonics}}  (\bibinfo{year}{2021}).

\bibitem{dai2020plasmonic}
\bibinfo{author}{Dai, Y.} \emph{et~al.}
\newblock \bibinfo{title}{Plasmonic topological quasiparticle on the nanometre
  and femtosecond scales}.
\newblock \emph{\bibinfo{journal}{Nature}} \textbf{\bibinfo{volume}{588}},
  \bibinfo{pages}{616--619} (\bibinfo{year}{2020}).

\bibitem{xiong2021polaritonic}
\bibinfo{author}{Xiong, L.} \emph{et~al.}
\newblock \bibinfo{title}{Polaritonic vortices with a half-integer charge}.
\newblock \emph{\bibinfo{journal}{Nano Letters}} \textbf{\bibinfo{volume}{21}},
  \bibinfo{pages}{9256--9261} (\bibinfo{year}{2021}).

\bibitem{ghosh2021topological}
\bibinfo{author}{Ghosh, A.} \emph{et~al.}
\newblock \bibinfo{title}{A topological lattice of plasmonic merons}.
\newblock \emph{\bibinfo{journal}{Applied Physics Reviews}}
  \textbf{\bibinfo{volume}{8}}, \bibinfo{pages}{041413} (\bibinfo{year}{2021}).

\bibitem{zhang2022optical}
\bibinfo{author}{Zhang, Q.} \emph{et~al.}
\newblock \bibinfo{title}{Optical topological lattices of bloch-type skyrmion
  and meron topologies}.
\newblock \emph{\bibinfo{journal}{Photonics Research}}
  \textbf{\bibinfo{volume}{10}}, \bibinfo{pages}{947--957}
  (\bibinfo{year}{2022}).

\bibitem{ghosh2023spin}
\bibinfo{author}{Ghosh, A.}, \bibinfo{author}{Yang, S.}, \bibinfo{author}{Dai,
  Y.} \& \bibinfo{author}{Petek, H.}
\newblock \bibinfo{title}{The spin texture topology of polygonal plasmon
  fields}.
\newblock \emph{\bibinfo{journal}{ACS Photonics}}  (\bibinfo{year}{2023}).

\bibitem{zhang2021bloch}
\bibinfo{author}{Zhang, Q.}, \bibinfo{author}{Xie, Z.}, \bibinfo{author}{Du,
  L.}, \bibinfo{author}{Shi, P.} \& \bibinfo{author}{Yuan, X.}
\newblock \bibinfo{title}{Bloch-type photonic skyrmions in optical chiral
  multilayers}.
\newblock \emph{\bibinfo{journal}{Physical Review Research}}
  \textbf{\bibinfo{volume}{3}}, \bibinfo{pages}{023109} (\bibinfo{year}{2021}).

\bibitem{krol2021observation}
\bibinfo{author}{Kr{\'o}l, M.} \emph{et~al.}
\newblock \bibinfo{title}{Observation of second-order meron polarization
  textures in optical microcavities}.
\newblock \emph{\bibinfo{journal}{Optica}} \textbf{\bibinfo{volume}{8}},
  \bibinfo{pages}{255--261} (\bibinfo{year}{2021}).

\bibitem{karnieli2021emulating}
\bibinfo{author}{Karnieli, A.}, \bibinfo{author}{Tsesses, S.},
  \bibinfo{author}{Bartal, G.} \& \bibinfo{author}{Arie, A.}
\newblock \bibinfo{title}{Emulating spin transport with nonlinear optics, from
  high-order skyrmions to the topological hall effect}.
\newblock \emph{\bibinfo{journal}{Nature Communications}}
  \textbf{\bibinfo{volume}{12}}, \bibinfo{pages}{1092} (\bibinfo{year}{2021}).

\bibitem{karnieli2022geometric}
\bibinfo{author}{Karnieli, A.}, \bibinfo{author}{Li, Y.} \&
  \bibinfo{author}{Arie, A.}
\newblock \bibinfo{title}{The geometric phase in nonlinear frequency
  conversion}.
\newblock \emph{\bibinfo{journal}{Frontiers of Physics}}
  \textbf{\bibinfo{volume}{17}}, \bibinfo{pages}{1--31} (\bibinfo{year}{2022}).

\bibitem{karnieli2018all}
\bibinfo{author}{Karnieli, A.} \& \bibinfo{author}{Arie, A.}
\newblock \bibinfo{title}{All-optical stern-gerlach effect}.
\newblock \emph{\bibinfo{journal}{Physical Review Letters}}
  \textbf{\bibinfo{volume}{120}}, \bibinfo{pages}{053901}
  (\bibinfo{year}{2018}).

\bibitem{jiang2017direct}
\bibinfo{author}{Jiang, W.} \emph{et~al.}
\newblock \bibinfo{title}{Direct observation of the skyrmion hall effect}.
\newblock \emph{\bibinfo{journal}{Nature Physics}}
  \textbf{\bibinfo{volume}{13}}, \bibinfo{pages}{162--169}
  (\bibinfo{year}{2017}).

\bibitem{chen2017skyrmion}
\bibinfo{author}{Chen, G.}
\newblock \bibinfo{title}{Skyrmion hall effect}.
\newblock \emph{\bibinfo{journal}{Nature Physics}}
  \textbf{\bibinfo{volume}{13}}, \bibinfo{pages}{112--113}
  (\bibinfo{year}{2017}).

\bibitem{gutierrez2021optical}
\bibinfo{author}{Guti{\'e}rrez-Cuevas, R.} \& \bibinfo{author}{Pisanty, E.}
\newblock \bibinfo{title}{Optical polarization skyrmionic fields in free
  space}.
\newblock \emph{\bibinfo{journal}{Journal of Optics}}
  \textbf{\bibinfo{volume}{23}}, \bibinfo{pages}{024004}
  (\bibinfo{year}{2021}).

\bibitem{liu2022disorder}
\bibinfo{author}{Liu, C.}, \bibinfo{author}{Zhang, S.}, \bibinfo{author}{Maier,
  S.~A.} \& \bibinfo{author}{Ren, H.}
\newblock \bibinfo{title}{Disorder-induced topological state transition in the
  optical skyrmion family}.
\newblock \emph{\bibinfo{journal}{Physical Review Letters}}
  \textbf{\bibinfo{volume}{129}}, \bibinfo{pages}{267401}
  (\bibinfo{year}{2022}).

\bibitem{gao2020paraxial}
\bibinfo{author}{Gao, S.} \emph{et~al.}
\newblock \bibinfo{title}{Paraxial skyrmionic beams}.
\newblock \emph{\bibinfo{journal}{Physical Review A}}
  \textbf{\bibinfo{volume}{102}}, \bibinfo{pages}{053513}
  (\bibinfo{year}{2020}).

\bibitem{gao2020e}
\bibinfo{author}{Gao, S.} \emph{et~al.}
\newblock \bibinfo{title}{Erratum: Paraxial skyrmionic beams}.
\newblock \emph{\bibinfo{journal}{Physical Review A}}
  \textbf{\bibinfo{volume}{104}}, \bibinfo{pages}{049901}
  (\bibinfo{year}{2021}).

\bibitem{beckley2010full}
\bibinfo{author}{Beckley, A.~M.}, \bibinfo{author}{Brown, T.~G.} \&
  \bibinfo{author}{Alonso, M.~A.}
\newblock \bibinfo{title}{Full poincar{\'e} beams}.
\newblock \emph{\bibinfo{journal}{Optics express}}
  \textbf{\bibinfo{volume}{18}}, \bibinfo{pages}{10777--10785}
  (\bibinfo{year}{2010}).

\bibitem{donati2016twist}
\bibinfo{author}{Donati, S.} \emph{et~al.}
\newblock \bibinfo{title}{Twist of generalized skyrmions and spin vortices in a
  polariton superfluid}.
\newblock \emph{\bibinfo{journal}{Proceedings of the National Academy of
  Sciences}} \textbf{\bibinfo{volume}{113}}, \bibinfo{pages}{14926--14931}
  (\bibinfo{year}{2016}).

\bibitem{shen2021topological}
\bibinfo{author}{Shen, Y.}
\newblock \bibinfo{title}{Topological bimeronic beams}.
\newblock \emph{\bibinfo{journal}{Optics Letters}}
  \textbf{\bibinfo{volume}{46}}, \bibinfo{pages}{3737--3740}
  (\bibinfo{year}{2021}).

\bibitem{shen2021generation}
\bibinfo{author}{Shen, Y.}, \bibinfo{author}{Mart{\'\i}nez, E.~C.} \&
  \bibinfo{author}{Rosales-Guzm{\'a}n, C.}
\newblock \bibinfo{title}{Generation of optical skyrmions with tunable
  topological textures}.
\newblock \emph{\bibinfo{journal}{ACS Photonics}} \textbf{\bibinfo{volume}{9}},
  \bibinfo{pages}{296--303} (\bibinfo{year}{2022}).

\bibitem{lin2021microcavity}
\bibinfo{author}{Lin, W.}, \bibinfo{author}{Ota, Y.}, \bibinfo{author}{Arakawa,
  Y.} \& \bibinfo{author}{Iwamoto, S.}
\newblock \bibinfo{title}{Microcavity-based generation of full poincar{\'e}
  beams with arbitrary skyrmion numbers}.
\newblock \emph{\bibinfo{journal}{Physical Review Research}}
  \textbf{\bibinfo{volume}{3}}, \bibinfo{pages}{023055} (\bibinfo{year}{2021}).

\bibitem{dennis2009singular}
\bibinfo{author}{Dennis, M.~R.}, \bibinfo{author}{O'holleran, K.} \&
  \bibinfo{author}{Padgett, M.~J.}
\newblock \bibinfo{title}{Singular optics: optical vortices and polarization
  singularities}.
\newblock In \emph{\bibinfo{booktitle}{Progress in optics}},
  vol.~\bibinfo{volume}{53}, \bibinfo{pages}{293--363}
  (\bibinfo{publisher}{Elsevier}, \bibinfo{year}{2009}).

\bibitem{nape2022revealing}
\bibinfo{author}{Nape, I.} \emph{et~al.}
\newblock \bibinfo{title}{Revealing the invariance of vectorial structured
  light in complex media}.
\newblock \emph{\bibinfo{journal}{Nature Photonics}}
  \textbf{\bibinfo{volume}{16}}, \bibinfo{pages}{538--546}
  (\bibinfo{year}{2022}).

\bibitem{klug2023robust}
\bibinfo{author}{Klug, A.}, \bibinfo{author}{Peters, C.} \&
  \bibinfo{author}{Forbes, A.}
\newblock \bibinfo{title}{Robust structured light in atmospheric turbulence}.
\newblock \emph{\bibinfo{journal}{Advanced Photonics}}
  \textbf{\bibinfo{volume}{5}}, \bibinfo{pages}{016006} (\bibinfo{year}{2023}).

\bibitem{shen2021supertoroidal}
\bibinfo{author}{Shen, Y.}, \bibinfo{author}{Hou, Y.},
  \bibinfo{author}{Papasimakis, N.} \& \bibinfo{author}{Zheludev, N.~I.}
\newblock \bibinfo{title}{Supertoroidal light pulses as electromagnetic
  skyrmions propagating in free space}.
\newblock \emph{\bibinfo{journal}{Nature communications}}
  \textbf{\bibinfo{volume}{12}}, \bibinfo{pages}{5891} (\bibinfo{year}{2021}).

\bibitem{guo2020meron}
\bibinfo{author}{Guo, C.}, \bibinfo{author}{Xiao, M.}, \bibinfo{author}{Guo,
  Y.}, \bibinfo{author}{Yuan, L.} \& \bibinfo{author}{Fan, S.}
\newblock \bibinfo{title}{Meron spin textures in momentum space}.
\newblock \emph{\bibinfo{journal}{Physical Review Letters}}
  \textbf{\bibinfo{volume}{124}}, \bibinfo{pages}{106103}
  (\bibinfo{year}{2020}).

\bibitem{sugic2021particle}
\bibinfo{author}{Sugic, D.} \emph{et~al.}
\newblock \bibinfo{title}{Particle-like topologies in light}.
\newblock \emph{\bibinfo{journal}{Nature communications}}
  \textbf{\bibinfo{volume}{12}}, \bibinfo{pages}{1--10} (\bibinfo{year}{2021}).

\bibitem{ornelas2022non}
\bibinfo{author}{Ornelas, P.}, \bibinfo{author}{Nape, I.},
  \bibinfo{author}{Koch, R. d.~M.} \& \bibinfo{author}{Forbes, A.}
\newblock \bibinfo{title}{Non-local skyrmions as topologically resilient
  quantum entangled states of light}.
\newblock \emph{\bibinfo{journal}{arXiv preprint arXiv:2210.04690}}
  (\bibinfo{year}{2022}).

\bibitem{zdagkas2020space}
\bibinfo{author}{Zdagkas, A.}, \bibinfo{author}{Papasimakis, N.},
  \bibinfo{author}{Savinov, V.} \& \bibinfo{author}{Zheludev, N.~I.}
\newblock \bibinfo{title}{Space-time nonseparable pulses: Constructing
  isodiffracting donut pulses from plane waves and single-cycle pulses}.
\newblock \emph{\bibinfo{journal}{Physical Review A}}
  \textbf{\bibinfo{volume}{102}}, \bibinfo{pages}{063512}
  (\bibinfo{year}{2020}).

\bibitem{hellwarth1996focused}
\bibinfo{author}{Hellwarth, R.} \& \bibinfo{author}{Nouchi, P.}
\newblock \bibinfo{title}{Focused one-cycle electromagnetic pulses}.
\newblock \emph{\bibinfo{journal}{Physical Review E}}
  \textbf{\bibinfo{volume}{54}}, \bibinfo{pages}{889} (\bibinfo{year}{1996}).

\bibitem{zdagkas2021observation}
\bibinfo{author}{Zdagkas, A.} \emph{et~al.}
\newblock \bibinfo{title}{Observation of toroidal pulses of light}.
\newblock \emph{\bibinfo{journal}{Nature Photonics}}
  \textbf{\bibinfo{volume}{16}}, \bibinfo{pages}{523--528}
  (\bibinfo{year}{2022}).

\bibitem{wan2022toroidal}
\bibinfo{author}{Wan, C.}, \bibinfo{author}{Cao, Q.}, \bibinfo{author}{Chen,
  J.}, \bibinfo{author}{Chong, A.} \& \bibinfo{author}{Zhan, Q.}
\newblock \bibinfo{title}{Toroidal vortices of light}.
\newblock \emph{\bibinfo{journal}{Nature Photonics}}
  \textbf{\bibinfo{volume}{16}}, \bibinfo{pages}{519--522}
  (\bibinfo{year}{2022}).

\bibitem{wan2022scalar}
\bibinfo{author}{Wan, C.}, \bibinfo{author}{Shen, Y.}, \bibinfo{author}{Chong,
  A.} \& \bibinfo{author}{Zhan, Q.}
\newblock \bibinfo{title}{Scalar optical hopfions}.
\newblock \emph{\bibinfo{journal}{eLight}} \textbf{\bibinfo{volume}{2}},
  \bibinfo{pages}{1--7} (\bibinfo{year}{2022}).

\bibitem{shen2022nondiffracting}
\bibinfo{author}{Shen, Y.}, \bibinfo{author}{Papasimakis, N.} \&
  \bibinfo{author}{Zheludev, N.~I.}
\newblock \bibinfo{title}{Nondiffracting supertoroidal pulses: Optical"
  k$\backslash$'arm$\backslash$'an vortex streets"}.
\newblock \emph{\bibinfo{journal}{arXiv preprint arXiv:2204.05676}}
  (\bibinfo{year}{2022}).

\bibitem{guo2021structured}
\bibinfo{author}{Guo, C.}, \bibinfo{author}{Xiao, M.},
  \bibinfo{author}{Orenstein, M.} \& \bibinfo{author}{Fan, S.}
\newblock \bibinfo{title}{Structured 3d linear space--time light bullets by
  nonlocal nanophotonics}.
\newblock \emph{\bibinfo{journal}{Light: Science \& Applications}}
  \textbf{\bibinfo{volume}{10}}, \bibinfo{pages}{1--15} (\bibinfo{year}{2021}).

\bibitem{nagaosa2012gauge}
\bibinfo{author}{Nagaosa, N.}, \bibinfo{author}{Yu, X.} \&
  \bibinfo{author}{Tokura, Y.}
\newblock \bibinfo{title}{Gauge fields in real and momentum spaces in magnets:
  monopoles and skyrmions}.
\newblock \emph{\bibinfo{journal}{Philosophical Transactions of the Royal
  Society A: Mathematical, Physical and Engineering Sciences}}
  \textbf{\bibinfo{volume}{370}}, \bibinfo{pages}{5806--5819}
  (\bibinfo{year}{2012}).

\bibitem{loder2017momentum}
\bibinfo{author}{Loder, F.}, \bibinfo{author}{Kampf, A.~P.},
  \bibinfo{author}{Kopp, T.} \& \bibinfo{author}{Braak, D.}
\newblock \bibinfo{title}{Momentum-space spin texture in a topological
  superconductor}.
\newblock \emph{\bibinfo{journal}{Physical Review B}}
  \textbf{\bibinfo{volume}{96}}, \bibinfo{pages}{024508}
  (\bibinfo{year}{2017}).

\bibitem{van2019photonic}
\bibinfo{author}{Van~Mechelen, T.} \& \bibinfo{author}{Jacob, Z.}
\newblock \bibinfo{title}{Photonic dirac monopoles and skyrmions: spin-1
  quantization}.
\newblock \emph{\bibinfo{journal}{Optical Materials Express}}
  \textbf{\bibinfo{volume}{9}}, \bibinfo{pages}{95--111}
  (\bibinfo{year}{2019}).

\bibitem{cisowski2023building}
\bibinfo{author}{Cisowski, C.}, \bibinfo{author}{Ross, C.} \&
  \bibinfo{author}{Franke-Arnold, S.}
\newblock \bibinfo{title}{Building paraxial optical skyrmions using rational
  maps}.
\newblock \emph{\bibinfo{journal}{Advanced Photonics Research}}
  \bibinfo{pages}{2200350} (\bibinfo{year}{2023}).

\bibitem{shen2023topologically}
\bibinfo{author}{Shen, Y.} \emph{et~al.}
\newblock \bibinfo{title}{Topologically controlled multiskyrmions in photonic
  gradient-index lenses}.
\newblock \emph{\bibinfo{journal}{arXiv preprint arXiv:2304.06332}}
  (\bibinfo{year}{2023}).

\bibitem{kuratsuji2021evolution}
\bibinfo{author}{Kuratsuji, H.} \& \bibinfo{author}{Tsuchida, S.}
\newblock \bibinfo{title}{Evolution of the stokes parameters, polarization
  singularities, and optical skyrmion}.
\newblock \emph{\bibinfo{journal}{Physical Review A}}
  \textbf{\bibinfo{volume}{103}}, \bibinfo{pages}{023514}
  (\bibinfo{year}{2021}).

\bibitem{shen2023topological}
\bibinfo{author}{Shen, Y.} \emph{et~al.}
\newblock \bibinfo{title}{Topological transformation and free-space transport
  of photonic hopfions}.
\newblock \emph{\bibinfo{journal}{Advanced Photonics}}
  \textbf{\bibinfo{volume}{5}}, \bibinfo{pages}{015001} (\bibinfo{year}{2023}).

\bibitem{zheng2018experimental}
\bibinfo{author}{Zheng, F.} \emph{et~al.}
\newblock \bibinfo{title}{Experimental observation of chiral magnetic bobbers
  in b20-type fege}.
\newblock \emph{\bibinfo{journal}{Nature Nanotechnology}}
  \textbf{\bibinfo{volume}{13}}, \bibinfo{pages}{451--455}
  (\bibinfo{year}{2018}).

\bibitem{tang2021magnetic}
\bibinfo{author}{Tang, J.} \emph{et~al.}
\newblock \bibinfo{title}{Magnetic skyrmion bundles and their current-driven
  dynamics}.
\newblock \emph{\bibinfo{journal}{Nature Nanotechnology}} \bibinfo{pages}{1--6}
  (\bibinfo{year}{2021}).

\bibitem{tai2019three}
\bibinfo{author}{Tai, J.-S.~B.} \& \bibinfo{author}{Smalyukh, I.~I.}
\newblock \bibinfo{title}{Three-dimensional crystals of adaptive knots}.
\newblock \emph{\bibinfo{journal}{Science}} \textbf{\bibinfo{volume}{365}},
  \bibinfo{pages}{1449--1453} (\bibinfo{year}{2019}).

\bibitem{ackerman2017diversity}
\bibinfo{author}{Ackerman, P.~J.} \& \bibinfo{author}{Smalyukh, I.~I.}
\newblock \bibinfo{title}{Diversity of knot solitons in liquid crystals
  manifested by linking of preimages in torons and hopfions}.
\newblock \emph{\bibinfo{journal}{Physical Review X}}
  \textbf{\bibinfo{volume}{7}}, \bibinfo{pages}{011006} (\bibinfo{year}{2017}).

\bibitem{yang2023first}
\bibinfo{author}{Yang, H.}, \bibinfo{author}{Liang, J.} \&
  \bibinfo{author}{Cui, Q.}
\newblock \bibinfo{title}{First-principles calculations for
  dzyaloshinskii--moriya interaction}.
\newblock \emph{\bibinfo{journal}{Nature Reviews Physics}}
  \textbf{\bibinfo{volume}{5}}, \bibinfo{pages}{43--61} (\bibinfo{year}{2023}).

\bibitem{zhang2020skyrmion}
\bibinfo{author}{Zhang, X.} \emph{et~al.}
\newblock \bibinfo{title}{Skyrmion-electronics: writing, deleting, reading and
  processing magnetic skyrmions toward spintronic applications}.
\newblock \emph{\bibinfo{journal}{Journal of Physics: Condensed Matter}}
  \textbf{\bibinfo{volume}{32}}, \bibinfo{pages}{143001}
  (\bibinfo{year}{2020}).

\bibitem{vzelezny2018spin}
\bibinfo{author}{{\v{Z}}elezn{\`y}, J.}, \bibinfo{author}{Wadley, P.},
  \bibinfo{author}{Olejn{\'\i}k, K.}, \bibinfo{author}{Hoffmann, A.} \&
  \bibinfo{author}{Ohno, H.}
\newblock \bibinfo{title}{Spin transport and spin torque in antiferromagnetic
  devices}.
\newblock \emph{\bibinfo{journal}{Nature Physics}}
  \textbf{\bibinfo{volume}{14}}, \bibinfo{pages}{220--228}
  (\bibinfo{year}{2018}).

\bibitem{yang2021chiral}
\bibinfo{author}{Yang, S.-H.}, \bibinfo{author}{Naaman, R.},
  \bibinfo{author}{Paltiel, Y.} \& \bibinfo{author}{Parkin, S.~S.}
\newblock \bibinfo{title}{Chiral spintronics}.
\newblock \emph{\bibinfo{journal}{Nature Reviews Physics}}
  \textbf{\bibinfo{volume}{3}}, \bibinfo{pages}{328--343}
  (\bibinfo{year}{2021}).

\bibitem{sohn2019light}
\bibinfo{author}{Sohn, H.~R.}, \bibinfo{author}{Liu, C.~D.},
  \bibinfo{author}{Wang, Y.} \& \bibinfo{author}{Smalyukh, I.~I.}
\newblock \bibinfo{title}{Light-controlled skyrmions and torons as
  reconfigurable particles}.
\newblock \emph{\bibinfo{journal}{Optics Express}}
  \textbf{\bibinfo{volume}{27}}, \bibinfo{pages}{29055--29068}
  (\bibinfo{year}{2019}).

\bibitem{sohn2020optically}
\bibinfo{author}{Sohn, H.~R.}, \bibinfo{author}{Liu, C.~D.},
  \bibinfo{author}{Voinescu, R.}, \bibinfo{author}{Chen, Z.} \&
  \bibinfo{author}{Smalyukh, I.~I.}
\newblock \bibinfo{title}{Optically enriched and guided dynamics of active
  skyrmions}.
\newblock \emph{\bibinfo{journal}{Optics Express}}
  \textbf{\bibinfo{volume}{28}}, \bibinfo{pages}{6306--6319}
  (\bibinfo{year}{2020}).

\bibitem{fujita2017ultrafast}
\bibinfo{author}{Fujita, H.} \& \bibinfo{author}{Sato, M.}
\newblock \bibinfo{title}{Ultrafast generation of skyrmionic defects with
  vortex beams: Printing laser profiles on magnets}.
\newblock \emph{\bibinfo{journal}{Physical Review B}}
  \textbf{\bibinfo{volume}{95}}, \bibinfo{pages}{054421}
  (\bibinfo{year}{2017}).

\bibitem{yang2018photonic}
\bibinfo{author}{Yang, W.}, \bibinfo{author}{Yang, H.}, \bibinfo{author}{Cao,
  Y.} \& \bibinfo{author}{Yan, P.}
\newblock \bibinfo{title}{Photonic orbital angular momentum transfer and
  magnetic skyrmion rotation}.
\newblock \emph{\bibinfo{journal}{Optics express}}
  \textbf{\bibinfo{volume}{26}}, \bibinfo{pages}{8778--8790}
  (\bibinfo{year}{2018}).

\bibitem{hirosawa2022laser}
\bibinfo{author}{Hirosawa, T.}, \bibinfo{author}{Klinovaja, J.},
  \bibinfo{author}{Loss, D.} \& \bibinfo{author}{D{\'\i}az, S.~A.}
\newblock \bibinfo{title}{Laser-controlled real-and reciprocal-space topology
  in multiferroic insulators}.
\newblock \emph{\bibinfo{journal}{Physical Review Letters}}
  \textbf{\bibinfo{volume}{128}}, \bibinfo{pages}{037201}
  (\bibinfo{year}{2022}).

\bibitem{zhao2021dzyaloshinskii}
\bibinfo{author}{Zhao, H.~J.}, \bibinfo{author}{Chen, P.},
  \bibinfo{author}{Prosandeev, S.}, \bibinfo{author}{Artyukhin, S.} \&
  \bibinfo{author}{Bellaiche, L.}
\newblock \bibinfo{title}{Dzyaloshinskii--moriya-like interaction in
  ferroelectrics and antiferroelectrics}.
\newblock \emph{\bibinfo{journal}{Nature Materials}}
  \textbf{\bibinfo{volume}{20}}, \bibinfo{pages}{341--345}
  (\bibinfo{year}{2021}).

\bibitem{junquera2021dzyaloshinskii}
\bibinfo{author}{Junquera, J.}
\newblock \bibinfo{title}{Dzyaloshinskii--moriya interaction turns electric}.
\newblock \emph{\bibinfo{journal}{Nature Materials}}
  \textbf{\bibinfo{volume}{20}}, \bibinfo{pages}{291--292}
  (\bibinfo{year}{2021}).

\bibitem{shi}
\bibinfo{author}{Shi, P.}, \bibinfo{author}{Du, L.}, \bibinfo{author}{Li, M.}
  \& \bibinfo{author}{Yuan, X.}
\newblock \bibinfo{title}{Symmetry-protected photonic chiral spin textures by
  spin–orbit coupling}.
\newblock \emph{\bibinfo{journal}{Laser Photonics Rev.}}
  \textbf{\bibinfo{volume}{15}}, \bibinfo{pages}{2000554}
  (\bibinfo{year}{2021}).

\bibitem{wu2022conformal}
\bibinfo{author}{Wu, H.-J.} \emph{et~al.}
\newblock \bibinfo{title}{Conformal frequency conversion for arbitrary
  vectorial structured light}.
\newblock \emph{\bibinfo{journal}{Optica}} \textbf{\bibinfo{volume}{9}},
  \bibinfo{pages}{187--196} (\bibinfo{year}{2022}).

\bibitem{watzel2020topological}
\bibinfo{author}{W{\"a}tzel, J.} \& \bibinfo{author}{Berakdar, J.}
\newblock \bibinfo{title}{Topological light fields for highly non-linear charge
  quantum dynamics and high harmonic generation}.
\newblock \emph{\bibinfo{journal}{Optics Express}}
  \textbf{\bibinfo{volume}{28}}, \bibinfo{pages}{19469--19481}
  (\bibinfo{year}{2020}).

\bibitem{jiang2020twisted}
\bibinfo{author}{Jiang, Y.} \emph{et~al.}
\newblock \bibinfo{title}{Twisted magnon as a magnetic tweezer}.
\newblock \emph{\bibinfo{journal}{Physical Review Letters}}
  \textbf{\bibinfo{volume}{124}}, \bibinfo{pages}{217204}
  (\bibinfo{year}{2020}).

\bibitem{wang2020optical}
\bibinfo{author}{Wang, X.-G.} \emph{et~al.}
\newblock \bibinfo{title}{The optical tweezer of skyrmions}.
\newblock \emph{\bibinfo{journal}{npj Computational Materials}}
  \textbf{\bibinfo{volume}{6}}, \bibinfo{pages}{1--7} (\bibinfo{year}{2020}).

\bibitem{padgett2011tweezers}
\bibinfo{author}{Padgett, M.} \& \bibinfo{author}{Bowman, R.}
\newblock \bibinfo{title}{Tweezers with a twist}.
\newblock \emph{\bibinfo{journal}{Nature Photonics}}
  \textbf{\bibinfo{volume}{5}}, \bibinfo{pages}{343--348}
  (\bibinfo{year}{2011}).

\bibitem{poy2022interaction}
\bibinfo{author}{Poy, G.} \emph{et~al.}
\newblock \bibinfo{title}{Interaction and co-assembly of optical and
  topological solitons}.
\newblock \emph{\bibinfo{journal}{Nature Photonics}}
  \textbf{\bibinfo{volume}{16}}, \bibinfo{pages}{454--461}
  (\bibinfo{year}{2022}).

\bibitem{tengdin2021imaging}
\bibinfo{author}{Tengdin, P.} \emph{et~al.}
\newblock \bibinfo{title}{Imaging the controllable rotation of a skyrmion
  crystal driven by femtosecond laser pulses}.
\newblock \emph{\bibinfo{journal}{arXiv preprint arXiv:2110.04548}}
  (\bibinfo{year}{2021}).

\bibitem{li2022highly}
\bibinfo{author}{Li, X.} \emph{et~al.}
\newblock \bibinfo{title}{Highly sensitive and topologically robust multimode
  sensing on spoof plasmonic skyrmions}.
\newblock \emph{\bibinfo{journal}{Advanced Optical Materials}}
  \bibinfo{pages}{2200331} (\bibinfo{year}{2022}).

\bibitem{yang2023spin}
\bibinfo{author}{Yang, A.} \emph{et~al.}
\newblock \bibinfo{title}{Spin-manipulated photonic skyrmion-pair for
  pico-metric displacement sensing}.
\newblock \emph{\bibinfo{journal}{Advanced Science}} \bibinfo{pages}{2205249}
  (\bibinfo{year}{2023}).

\bibitem{yuan2019detecting}
\bibinfo{author}{Yuan, G.~H.} \& \bibinfo{author}{Zheludev, N.~I.}
\newblock \bibinfo{title}{Detecting nanometric displacements with optical ruler
  metrology}.
\newblock \emph{\bibinfo{journal}{Science}} \textbf{\bibinfo{volume}{364}},
  \bibinfo{pages}{771--775} (\bibinfo{year}{2019}).

\bibitem{du2021ultrasensitive}
\bibinfo{author}{Du, L.}, \bibinfo{author}{Yang, A.} \& \bibinfo{author}{Yuan,
  X.}
\newblock \bibinfo{title}{Ultrasensitive displacement sensing method and device
  based on local spin characteristics} (\bibinfo{year}{2021}).
\newblock \bibinfo{note}{US Patent App. 16/303,773}.

\bibitem{shen2019optical}
\bibinfo{author}{Shen, Y.} \emph{et~al.}
\newblock \bibinfo{title}{Optical vortices 30 years on: Oam manipulation from
  topological charge to multiple singularities}.
\newblock \emph{\bibinfo{journal}{Light: Science \& Applications}}
  \textbf{\bibinfo{volume}{8}}, \bibinfo{pages}{1--29} (\bibinfo{year}{2019}).

\bibitem{galiffi2022photonics}
\bibinfo{author}{Galiffi, E.} \emph{et~al.}
\newblock \bibinfo{title}{Photonics of time-varying media}.
\newblock \emph{\bibinfo{journal}{Advanced Photonics}}
  \textbf{\bibinfo{volume}{4}}, \bibinfo{pages}{014002} (\bibinfo{year}{2022}).

\bibitem{ge2021observation}
\bibinfo{author}{Ge, H.} \emph{et~al.}
\newblock \bibinfo{title}{Observation of acoustic skyrmions}.
\newblock \emph{\bibinfo{journal}{Physical Review Letters}}
  \textbf{\bibinfo{volume}{127}}, \bibinfo{pages}{144502}
  (\bibinfo{year}{2021}).

\bibitem{muelas2022observation}
\bibinfo{author}{Muelas-Hurtado, R.~D.} \emph{et~al.}
\newblock \bibinfo{title}{Observation of polarization singularities and
  topological textures in sound waves}.
\newblock \emph{\bibinfo{journal}{Physical Review Letters}}
  \textbf{\bibinfo{volume}{129}}, \bibinfo{pages}{204301}
  (\bibinfo{year}{2022}).

\bibitem{hu2023observation}
\bibinfo{author}{Hu, P.} \emph{et~al.}
\newblock \bibinfo{title}{Observation of localized acoustic skyrmions}.
\newblock \emph{\bibinfo{journal}{Applied Physics Letters}}
  \textbf{\bibinfo{volume}{122}}, \bibinfo{pages}{022201}
  (\bibinfo{year}{2023}).

\bibitem{cao2023observation}
\bibinfo{author}{Cao, L.}, \bibinfo{author}{Wan, S.}, \bibinfo{author}{Zeng,
  Y.}, \bibinfo{author}{Zhu, Y.} \& \bibinfo{author}{Assouar, B.}
\newblock \bibinfo{title}{Observation of phononic skyrmions based on hybrid
  spin of elastic waves}.
\newblock \emph{\bibinfo{journal}{Science Advances}}
  \textbf{\bibinfo{volume}{9}}, \bibinfo{pages}{eadf3652}
  (\bibinfo{year}{2023}).

\bibitem{parmee2022optical}
\bibinfo{author}{Parmee, C.~D.}, \bibinfo{author}{Dennis, M.~R.} \&
  \bibinfo{author}{Ruostekoski, J.}
\newblock \bibinfo{title}{Optical excitations of skyrmions, knotted solitons,
  and defects in atoms}.
\newblock \emph{\bibinfo{journal}{Communications Physics}}
  \textbf{\bibinfo{volume}{5}}, \bibinfo{pages}{1--8} (\bibinfo{year}{2022}).

\end{thebibliography}

\end{document}